\documentclass[12pt,letter]{article}

\usepackage{graphicx, epsfig, color}
\def\eslt{E_T^{\rm miss}}
\def\delew{\Delta_{\rm EW}}

\def\to{\rightarrow}

\def\bi{\begin{itemize}}
\def\ei{\end{itemize}}
\def\te{\tilde e}

\def\tu{\tilde u}
\def\sps1ap{SPS1a$^\prime$}
\def\c1p{C1$^\prime$}

\def\tb{\tilde b}

\def\tst{\tilde t}
\def\ttau{\tilde \tau}

\def\tg{\tilde g}
\def\tnu{\tilde\nu}

\def\tq{\tilde q}
\def\tw{\widetilde W}
\def\tz{\widetilde Z}
\def\alt{\stackrel{<}{\sim}}
\def\agt{\stackrel{>}{\sim}}
\def\be{\begin{equation}}  
\def\ee{\end{equation}}  
\def\bea{\begin{eqnarray}}  
\def\eea{\end{eqnarray}}  
\def\beas{\begin{eqnarray*}}  
\def\eeas{\end{eqnarray*}}  
\newcommand\prd[3]{{\it Phys.\ Rev.\ }{\bf D #1} (#2) #3}

\newcommand\prl[3]{{\it Phys.\ Rev.\ Lett.\ }{\bf #1} (#2) #3}
\newcommand\plb[3]{{\it Phys.\ Lett.\ }{\bf B #1} (#2) #3}
\newcommand\jhep[3]{{\it J. High Energy Phys.\ }{\bf #1} (#2) #3}

\newcommand\npb[3]{{\it Nucl.\ Phys.\ }{\bf B #1} (#2) #3}
\newcommand\epjc[3]{{\it Eur.\ Phys.\ J. }{\bf C #1} (#2) #3}

\newcommand{\hepph}[1]{hep-ph/#1}

\begin{document}
\begin{titlepage}
\begin{flushright}
UH-511-1242-15
\end{flushright}

\vspace{0.5cm}
\begin{center}
{\Large \bf Natural SUSY with a bino- or wino-like LSP
}\\ 
\vspace{1.2cm} \renewcommand{\thefootnote}{\fnsymbol{footnote}}
{\large Howard Baer$^{1,2}$\footnote[1]{Email: baer@nhn.ou.edu },
Vernon Barger$^3$\footnote[2]{Email: barger@pheno.wisc.edu },
Peisi Huang$^{4,5}$\footnote[3]{Email: peisi@uchicago.edu},\\
Dan Mickelson$^1$\footnote[3]{Email: dsmickelson@ou.edu},
Maren Padeffke-Kirkland$^1$\footnote[4]{Email: m.padeffke@ou.edu } 
and Xerxes Tata$^6$\footnote[5]{Email: tata@phys.hawaii.edu}
}\\ 
\vspace{1.2cm} \renewcommand{\thefootnote}{\arabic{footnote}}
{\it 
$^1$Dept. of Physics and Astronomy,
University of Oklahoma, Norman, OK 73019, USA \\
}
{\it 
$^2$William I. Fine Theoretical Physics Institute,
University of Minnesota, Minneapolis, MN 55455, USA \\
}
{\it 
$^3$Dept. of Physics,
University of Wisconsin, Madison, WI 53706, USA \\
}
{\it 
$^4$ HEP Division, Argonne National Lab, Argonne, IL, 60439, USA\\
}
{\it 
$^5$ Enrico Fermi Institute, University of Chicago, Chicago, IL 60637, USA\\
}
{\it 
$^6$Dept. of Physics and Astronomy,
University of Hawaii, Honolulu, HI 96822, USA \\
}

\end{center}

\vspace{0.5cm}
\begin{abstract}
\noindent 
In natural SUSY models higgsinos are always light because
$\mu^2$ cannot be much larger than
$M_Z^2$, while squarks and gluinos may be
very heavy.  Unless gluinos are discovered at LHC13, the commonly
assumed unification of gaugino mass parameters will imply correspondingly
heavy winos and binos, resulting in a higgsino-like LSP and small
inter-higgsino mass splittings. The small visible energy release in
higgsino decays makes 
their
pair production difficult to detect at the LHC.
Relaxing gaugino mass universality allows for relatively
light winos and binos
without violating LHC gluino
mass bounds and without affecting naturalness.  
In the case where the bino mass $M_1\alt \mu$,
then one obtains a mixed bino-higgsino LSP with instead sizable
$\tw_1-\tz_1$ and $\tz_2-\tz_1$ mass gaps.  The thermal neutralino
abundance can match the measured dark matter density in contrast to
models with a higgsino-like LSP where WIMPs 
(weakly interacting massive particles) are underproduced by
factors of 10-15.  If instead $M_2\alt \mu$, then one obtains a mixed
wino-higgsino LSP with large $\tz_2-\tz_1$ but small $\tw_1-\tz_1$
mass gaps with still an under-abundance of thermally-produced
WIMPs. 
%
We discuss dark matter
detection in other direct and indirect detection experiments and
caution that the bounds from these must be interpreted with care. Finally, we
show that LHC13 experiments should be able to probe these
non-universal mass scenarios via a variety of channels including
multi-lepton + $\eslt$ events, $WZ+\eslt$ events, $Wh+\eslt$ events
and $W^\pm W^\pm +\eslt$ events from electroweak chargino and
neutralino production.
\vspace*{0.8cm}

\end{abstract}

\end{titlepage}

\section{Introduction}

Results from the first extended runs of LHC at $\sqrt{s}=7$ and 8 TeV
have led some authors to imply that there is a crisis in supersymmetry (SUSY)
phenomenology \cite{Lykken:2014bca}: how can it be that the Higgs and
vector boson masses -- whose values are related to weak scale soft SUSY
breaking (SSB) parameters and to the superpotential parameter $\mu$ -- are
clustered near 100 GeV while superpartner masses, whose values are also
determined by soft SUSY breaking terms, are so heavy that they are
beyond the reach of LHC?  The superpotential higgsino mass parameter
$\mu$ and the SSB Higgs mass parameters enter via the tree-level Higgs
potential, whereas other SSB parameters -- specifically, those that
affect sparticles with the largest couplings to the Higgs sector -- only
enter at higher order. This is clearly evident, for example, in the
well-known expression,
\be
\frac{M_Z^2}{2}=\frac{m_{H_d}^2+\Sigma_d^d-(m_{H_u}^2+\Sigma_u^u)\tan^2\beta}{\tan^2\beta -1}-\mu^2\;,
\label{eq:mzs}
\ee
for the $Z$ mass, where $\Sigma_u^u$ and $\Sigma_d^d$ denote the 1-loop
corrections explicitly given in the Appendix of Ref.~\cite{rns}. SUSY
models requiring large cancellations between the various terms on the
right-hand-side of (\ref{eq:mzs}) to reproduce the measured value of
$M_Z^2$ are regarded as unnatural, or fine-tuned.\footnote{We emphasize
that for superpartners up to a few TeV range, the degree of
fine-tuning that we are talking about is many orders of magnitude smaller than
in the Standard Model because  scalar masses do not
exhibit quadratic sensitivity to physics at the ultra-high scale
if SUSY is softly broken.}

Several measures have been proposed \cite{eenz,bg,kn,ltr,rns} to
quantify the degree of fine-tuning. A common feature of these is that
they all regard the model to be fine-tuned if $\mu^2 \gg M_Z^2$. This is
because in most models $\mu$ directly enters Eq.~(\ref{eq:mzs}) as an
{\em independent parameter}, and unexplained cancellations have then to
be invoked to obtain the observed value of $M_Z$. In contrast, in most
models the SSB masses are obtained in terms of a one (or more)
model-parameters and so are not independent, allowing for the
possibility of large cancellations that is ignored by the
commonly used large log measure. It is a neglect of these
parameter-correlations that has led some authors to conclude that light
top-squarks are a necessary feature of natural SUSY.  In fact, as we
have just argued, it is $|\mu| \sim M_Z$ and concomitantly the existence
of light higgsinos\footnote{We assume here that there is no
SUSY-breaking higgsino mass term (such a term would lead to hard SUSY
breaking -- and so would be automatically forbidden -- if higgsinos had
superpotential Yukawa couplings to any Standard Model singlet
\cite{girardello}) so that the higgsino mass comes only from the
superpotential parameter $\mu$. This is the case in all models that we
know of.}  (and not light stops) that is the robust conclusion of naturalness
considerations. The importance of low $\mu$ for electroweak naturalness 
was recognized by Chan, Chattopadhyay and Nath \cite{ccn} over fifteen
years ago and has recently been emphasized in 
Refs. \cite{ltr,rns,vp} and by Martin \cite{seminat}.

In earlier papers we have developed the radiatively-driven natural
SUSY (RNS) framework characterized by values of the parameter $\delew =
10-30$ range corresponding to 3-10\% electroweak fine-tuning
\cite{ltr,rns,vp}. Within this framework,
\bi
\item the superpotential $\mu$ term has magnitude $|\mu|\sim 100-300$ GeV 
(the closer to $M_Z$ the better);

\item the up-Higgs soft term $m_{H_u}^2$ is driven radiatively to small
negative values $m_{H_u}^2(weak)\sim -M_Z^2$;

\item the magnitude of radiative corrections contained in $\Sigma_u^u$ 
should be smaller than or comparable to $M_Z^2$.  
This latter condition occurs for TeV-scale
highly mixed top squarks- a situation which also lifts $m_h$ into the
125 GeV regime \cite{ltr}.  In contrast, the terms $m_{H_d}^2$ and
$\Sigma_d^d$ can occur at the multi-TeV level since they are suppressed
by $\tan^2\beta$ where $\tan\beta$ is required to be in the 3-50 range.

\item Since the gluino mass feeds into the stop masses via RG evolution--
  and thus into $\Sigma_u^u(\tst_{1,2})$-- then low $\Delta_{\rm EW}$ also
  requires an upper bound on $m_{\tg} \alt 4-5$ TeV \cite{rns}. 
  Of course, $M_3$ is also bounded from below by the experimental bound 
  of $m_{\tg}\agt 1.3$~TeV based on LHC8 searches within the context
  of SUSY models like mSUGRA/CMSSM \cite{atlas_susy} or within simplified
  models \cite{cms_susy}.

\item First and second generation
  sfermions can be allowed anywhere in the $\sim 5-20$~TeV range 
  without jeapordizing naturalness \cite{maren}. 
  The higher range of values ameliorates  the SUSY flavour, $CP$, 
  gravitino and proton-decay problems due to decoupling.
\ei
Inspired by gauge coupling unification, in these previous studies we
had assumed gaugino mass unification as well as naturalness. 
From gaugino mass unification one expects at the weak scale that 
$M_1\sim M_3/7$ and $M_2\sim 2M_3/7$ so that the LHC8 lower bound on $M_3$
also provides a lower bound on $M_1$ and $M_2$. 
In this case, for natural SUSY which respects LHC8 bounds, 
we expect the mass hierarchy $|\mu| < M_1 < M_2 < M_3$ to occur.  
Thus, in the RNS model which we take as the paradigm case for the study of natural
SUSY, one expects four light higgsino states with mass
$m_{\tw_1^{\pm}},\ m_{\tz_{1,2}}\sim |\mu|$ where the lightest higgsino
$\tz_1$ acts as the lightest-SUSY-particle or LSP. In particular,
mixed higgsino-bino or higgino-wino LSPs are not allowed if the gluino is heavy.

Collider signals as well as cosmology depend sensitively on the nature
of the LSP. For instance, in the RNS framework with gaugino masses near the TeV
range, we expect the light electroweak -inos $\tw_1^\pm$ and $\tz_{1,2}$ to be
dominantly higgsino-like with typically small $m_{\tw_1}-m_{\tz_1}$ and
$m_{\tz_2}-m_{\tz_1}$ mass splittings of order 10-20 GeV \cite{rns}.
Such a small mass splitting results in only soft visible energy release
from the heavier higgsino three-body decays to the $\tz_1$.  This
situation makes pair production of higgsinos very difficult to detect at
LHC \cite{rns@lhc,chan,monojet,kribs,dilep} in spite of their relatively
small masses and correspondingly large production cross sections; other
superpartners may be very heavy, and possibly beyond the reach of the
LHC. In contrast, in models with light gauginos and heavy higgsinos, the
mass gap between the bino and wino-like states tends to be large (if
gaugino mass unification is assumed), and signals from wino pair
production followed by their decays to bino-like LSPs should be readily
detectable. The celebrated clean trilepton signature arising from $\tw_1\tz_2$ 
production is perhaps the best-known example.

The phenomenology of dark matter is even much more sensitive to the
content of the LSP.  Higgsino and wino-like LSPs lead to an
under-abundance of thermally-produced LSPs whereas a bino-like LSP leads
to overproduction of WIMPs (weakly interacting massive particles) unless
the neutralino annihilation rate is dynamically enhanced, {\it e.g.} via
an $s$-channel resonance or via co-annihilation, or their density is
diluted by entropy production late in the history of the Universe.  In
the wino- or higgsino-LSP cases, if one solves the strong $CP$ problem
via a quasi-visible axion \cite{axion}, then the dark matter is expected
to occur as an axion-neutralino admixture, {\it i.e.} two dark matter
particles \cite{mixedDM}.

Gaugino mass unification -- well-motivated as it may be -- is by no means
sacrosanct. Phenomenologically, while the high scale value of $M_3$ is
required to be large by LHC8 constraints on $m_{\tg}$, $M_1$ and/or
$M_2$ may well have much smaller magnitudes without impacting naturalness.
These considerations motivated us to examine how the phenomenology of
natural SUSY models with $|\mu| \sim 100-300$~GeV may be altered if we
give up the gaugino mass unification assumption and allow for the
possibility that the bino or/and wino also happens to be light.  The LSP
(and possibly also other electroweak-inos) would then be mixtures of
higgsinos and electroweak gauginos, or may even be very nearly bino- or
wino-like, resulting in very different mass and mixing patterns from
expectations within the RNS framework.  A mixed bino-higgsino LSP could
well lead to the observed relic-density for thermally produced
neutralinos. We acknowledge that small values of gaugino mass parameters
would have to be regarded as fortuitous from the perspective of
naturalness. Nevertheless since light winos/binos do not jeapordize
naturalness, in the absence of any compelling theory of the origin of
SSB parameters, we felt a phenomenological study of this situation is
justified by our philosophy that it is best to ``leave no stone
unturned'' in the search for natural SUSY at the LHC. 

Non-universal gaugino masses (NUGM) can occur in GUT models wherein the
gauge kinetic function transforms non-trivially as the direct product of
two adjoints \cite{gkf,anderson}.  Or, it may be that GUTs play no role, and
that unification occurs within the string-model context. Models with
mixed anomaly- and gravity-mediation contributions to gauginos masses
also lead to non-universal gaugino mass parameters \cite{kklt}.  
Investigation of how the phenomenology of natural SUSY models is
modified from RNS expectations forms the subject of this paper. 
Naturalness in the context of non-universal gaugino masses has also been
considered in Refs. \cite{shafi} and \cite{seminat}.

\subsection{Natural SUSY benchmark scenarios}

We begin by exhibiting a sample benchmark point within the framework of
the canonical 2-extra-parameter non-universal Higgs model (NUHM2) with
{\it unified} gaugino mass parameters and a higgsino-like LSP under the
column RNSh in Table \ref{tab:bm}.  This point has parameters $m_0=5000$
GeV, $m_{1/2}=700$ GeV, $A_0=-8000$ GeV and $\tan\beta =10$ with $(\mu
,\ m_A)=(200,1000)$ GeV.  The RNSh point has $\Delta_{\rm EW}=9.6$
corresponding to about 10\% electroweak fine-tuning, and $m_h=124.3$ GeV
while $m_{\tg}\simeq 1.8$ TeV with $m_{\tq}=5.2$ TeV.  It is safely
beyond LHC8 reach. The lightest neutralino is dominantly higgsino-like
(higgsino-wino-bino composition is listed as $v_h^{(1)} \equiv
\sqrt{v_1^{{(1)2}}+v_2^{{(1)2}}}$, $v_w^{(1)}$ and $v_b^{(1)}$ defined
similarly to Ref. \cite{wss}) and has mass $m_{\tz_1}=188$ GeV and
thermally-produced neutralino relic density \cite{isared}
$\Omega_{\tz_1}h^2=0.013$.  SUSY contributions to the branching fraction
for $b\to s\gamma$ are negligible so that this is close to its SM value
\cite{bsgth} and in accord with experiment \cite{bsgamma}. The
spin-independent neutralino-proton scattering cross section shown in the
third-last row of the table naively violates the bound
$\sigma^{SI}(\tz_1 p) \alt (2-3)\times 10^{-9}$~pb from the LUX
experiment \cite{lux}, but we note that this bound is obtained assuming
that the neutralino comprises {\it all} of the cold dark matter. In our
case, the thermal neutralino contribution is just about 10\% of the
total DM contribution, and this point is in accord with the constraint
upon scaling the expected event rate by $\xi =
\Omega_{\tz_1}h^2/0.12$. \footnote{We remark that other processes may
further alter the neutralino relic density from its thermal value,
increasing it if there are late decays of heavy particles to
neutralinos, or diluting it if these decay into SM particles. For more
detailed discussion of non-thermally-produced dark matter, see the
recent review \cite{review}.}  We also show the spin-dependent
neutralino-nucleon scattering cross-section.  The IceCube experiment
currently has the best sensitivity to this quantity by searching for
high energy neutrinos arising from neutralinos which are captured by the
sun and annihilated in the solar core.  The current IceCube limit
\cite{icecube}, lies around $\sigma^{SD}(\tz_1 p)\alt 1.5\times 10^{-4}$
pb so that the RNSh point would seem to be excluded by this bound.  For
this analysis, the neutralino density in the solar core is obtained by
assuming equilibration between the capture rate and the annihilation
rate of neutralinos.  Since the capture rate scales {\em linearly} with
the neutralino relic density, the predicted event rates also need to be
scaled by $\xi$ before comparing with IceCube.  After re-scaling, we see
that the RNSh point is an order of magnitude away from the IceCube upper
limit of $\sim 1.5\times 10^{-4}$~pb that is obtained assuming the
neutralinos dominantly annihilate via $\tz_1\tz_1 \to WW$.
%
%
The
other columns display natural SUSY benchmark points where the bino or
the wino mass parameters are dialed to relatively low values resulting
in natural SUSY models with either a bino-like (RNSb) or wino-like
(RNSw) LSP. These cases will be discussed in detail in the following
sections.

\subsection{Remainder of paper}

The remainder of this paper is organized as follows.  
In Sec. \ref{sec:bino}, we first investigate the case of $|M_1|\sim |\mu| \ll M_{2,3}$ 
where we treat $M_1$ as an additional phenomenological parameter.\footnote{We 
frequently denote both the GUT and weak scale values of the gaugino mass parameters by $M_i$. 
We assume that it will be clear from the context which case is being used 
so that this abuse of notation will not cause any confusion.} 
In this case, the LSP can become mixed bino-higgsino or
even mainly bino-like.  Note also that while we can always choose one of
the gaugino mass parameters to be positive, the signs of the remaining
ones are physical. In our study, we will examine both signs of the
gaugino mass parameters that are assumed to depart from universality. In
Sec. \ref{sec:wino}, we investigate the case with $|M_2|\sim |\mu|\ll
M_{1,3}$ which can generate a wino-like LSP.  In Sec. \ref{sec:gen}, we
examine the more general case where both $|M_1|$ and $|M_2|$ are
simultaneously comparable to $|\mu|$. Our conclusions are presented in
Sec. \ref{sec:conclude}.

%
\begin{table}\centering
\begin{tabular}{lccc}
\hline
parameter & RNS$h$ & RNS$b$ & RNS$w$ \\
\hline
$M_1$(GUT)      & 700 & 380    & 700 \\
$M_2$(GUT)      & 700 & 700 & 175 \\
$M_3$(GUT)     & 700 & 700 & 700 \\
\hline
$m_{\tg}$   & 1795.8  & 1796.2  & 1809.8  \\
$m_{\tu_L}$ & 5116.2  & 5116.2 &  5100.7 \\
$m_{\tu_R}$ & 5273.3  & 5271.3 & 5277.4 \\
$m_{\te_R}$ & 4809.0  & 4804.4 & 4806.7 \\
$m_{\tst_1}$& 1435.1  & 1438.1 &  1478.3 \\
$m_{\tst_2}$& 3601.2  & 3603.3 &  3584.9 \\
$m_{\tb_1}$ & 3629.4  & 3631.5 & 3611.6 \\
$m_{\tb_2}$ & 5003.9  & 5003.6 & 5007.4 \\
$m_{\ttau_1}$ & 4735.6& 4731.1 & 4733.9 \\
$m_{\ttau_2}$ & 5071.9& 5070.8 & 5053.9 \\
$m_{\tnu_{\tau}}$ & 5079.2 & 5078.1 & 5060.8 \\
$m_{\tw_2}$ & 610.9   & 611.0 & 248.4 \\
$m_{\tw_1}$ & 205.3   & 205.3  & 121.5\\
$m_{\tz_4}$ & 621.4 & 621.5  & 322.1 \\
$m_{\tz_3}$ & 322.0 & 217.9  & 237.8 \\
$m_{\tz_2}$ & 209.3 & 209.8 & 211.8 \\
$m_{\tz_1}$ & 187.8 & 149.5 & 114.2 \\
$m_h$       & 124.3 & 124.2 &  124.3 \\
\hline
$v_h^{(1)}$ & 0.96 & 0.57 & 0.60\\
$v_w^{(1)}$ & -0.14 & 0.07 & -0.80\\
$v_b^{(1)}$ & 0.24 & -0.82 & 0.08\\
\hline
$\Delta_{\rm EW}$ & 9.6 & 9.6 & 10.8 \\
$\Omega_{\tz_1}^{std}h^2$ & 0.013 & 0.11 & 0.0015 \\
$BF(b\to s\gamma)$ & $3.3\times 10^{-4}$  & $3.3\times 10^{-4}$ & $3.3\times 10^{-4}$ \\
$\sigma^{SI}(\tz_1 p)$ (pb) & $1.6\times 10^{-8}$  & $1.7\times 10^{-8}$ & $4.3\times 10^{-8}$\\
$\sigma^{SD}(\tz_1 p)$ (pb) & $1.7\times 10^{-4}$  & $2.8\times 10^{-4}$ & $8.9\times 10^{-4}$\\
$\langle\sigma v\rangle |_{v\to 0}$  (cm$^3$/sec)
& $2.0\times 10^{-25}$  & $1.8\times 10^{-26}$ & $1.7\times 10^{-24}$\\
\hline
\end{tabular}
\caption{Input parameters and masses in~GeV units for three Natural SUSY
benchmark points with $\mu =200$ GeV and $m_A=1000$ GeV. 
We also take $m_0=5000$ GeV, $A_0=-8000$ GeV and $\tan\beta =10$.
Also shown are the values of several non-accelerator observables.  }
\label{tab:bm}
\end{table}

\section{Natural SUSY with a bino-like LSP}
\label{sec:bino}

\begin{figure}[tbp]
\begin{center}
\includegraphics[height=0.3\textheight]{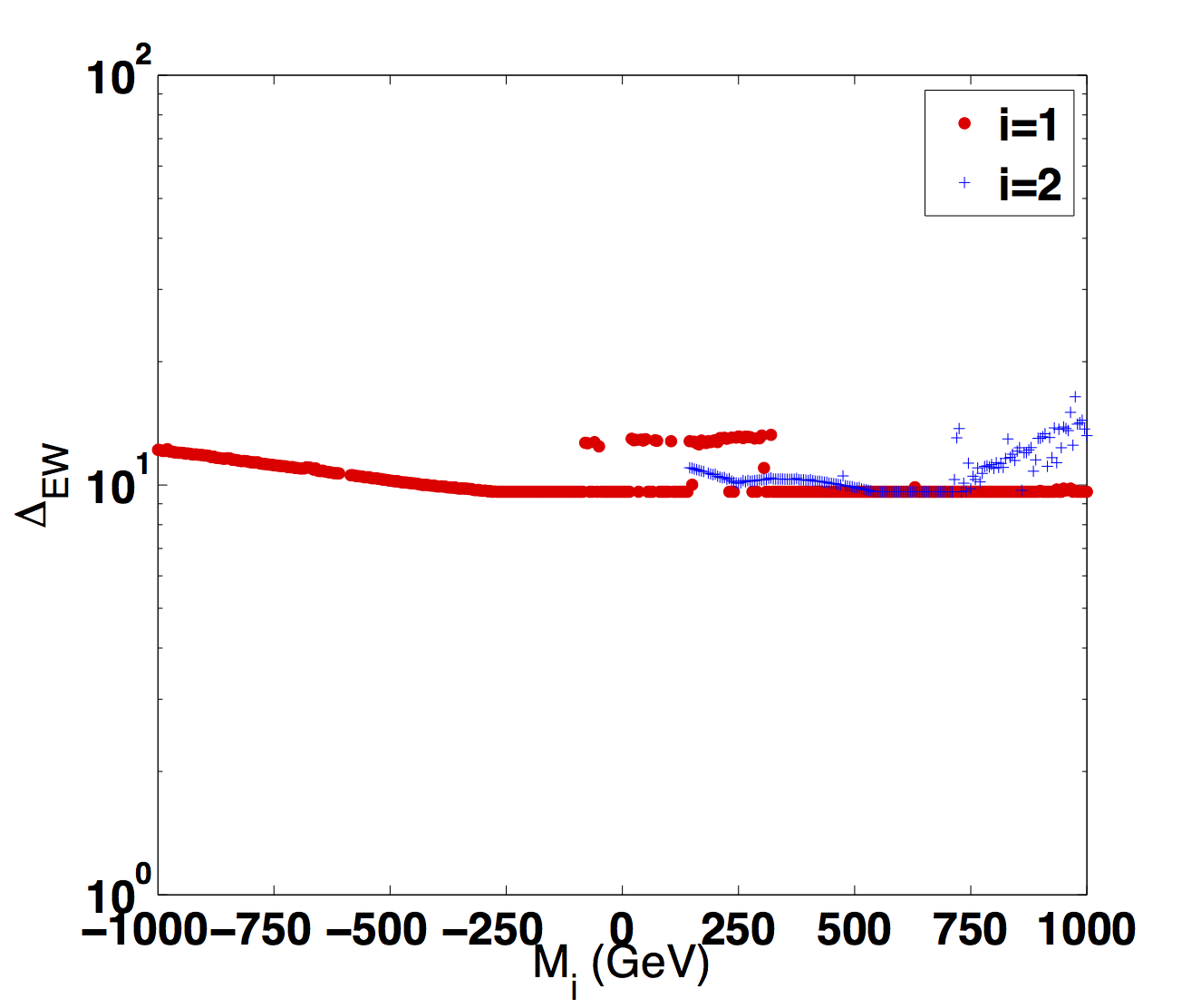}
\end{center}
\caption{Variation in fine-tuning measure $\Delta_{\rm EW}$ vs. $M_1$
  (red circles) or $M_2$ (blue pluses), with all other parameters fixed
at their values for the RNS SUSY benchmark model point in Table~\ref{tab:bm}.
Here, and in subsequent figures the $M_i$ on the horizontal axis is the 
value of the corresponding gaugino mass parameter renormalized at the
GUT scale.  We cut the graphs off if the lighter chargino mass falls
below 100~GeV.
\label{fig:M12_EW}}
\end{figure}

In this section, we examine how the phenomenology of natural SUSY models
is altered if we allow for non-universal gaugino mass parameters, and
let the GUT scale bino mass vary independently.  To this end, we adopt
the RNSh benchmark point from Table \ref{tab:bm}, but now allow $M_1$ to
be a free parameter, positive or negative.  To generate spectra and the
value of $\Delta_{\rm EW}$, we adopt the Isajet 7.84 spectrum generator
\cite{isajet}. In Fig. \ref{fig:M12_EW}, we show by red circles the
value of $\Delta_{\rm EW}$ versus the GUT scale value of $M_1$. 
We see that -- 
aside from numerical instabilities arising from our iterative solution
to the SUSY RGEs -- 
the value of $\Delta_{\rm EW}$ stays nearly constant 
so that, as anticipated, varying $M_1$ hardly affects the
degree of electro-weak fine-tuning.

\begin{figure}[tbp]
\begin{center}
\includegraphics[height=0.3\textheight]{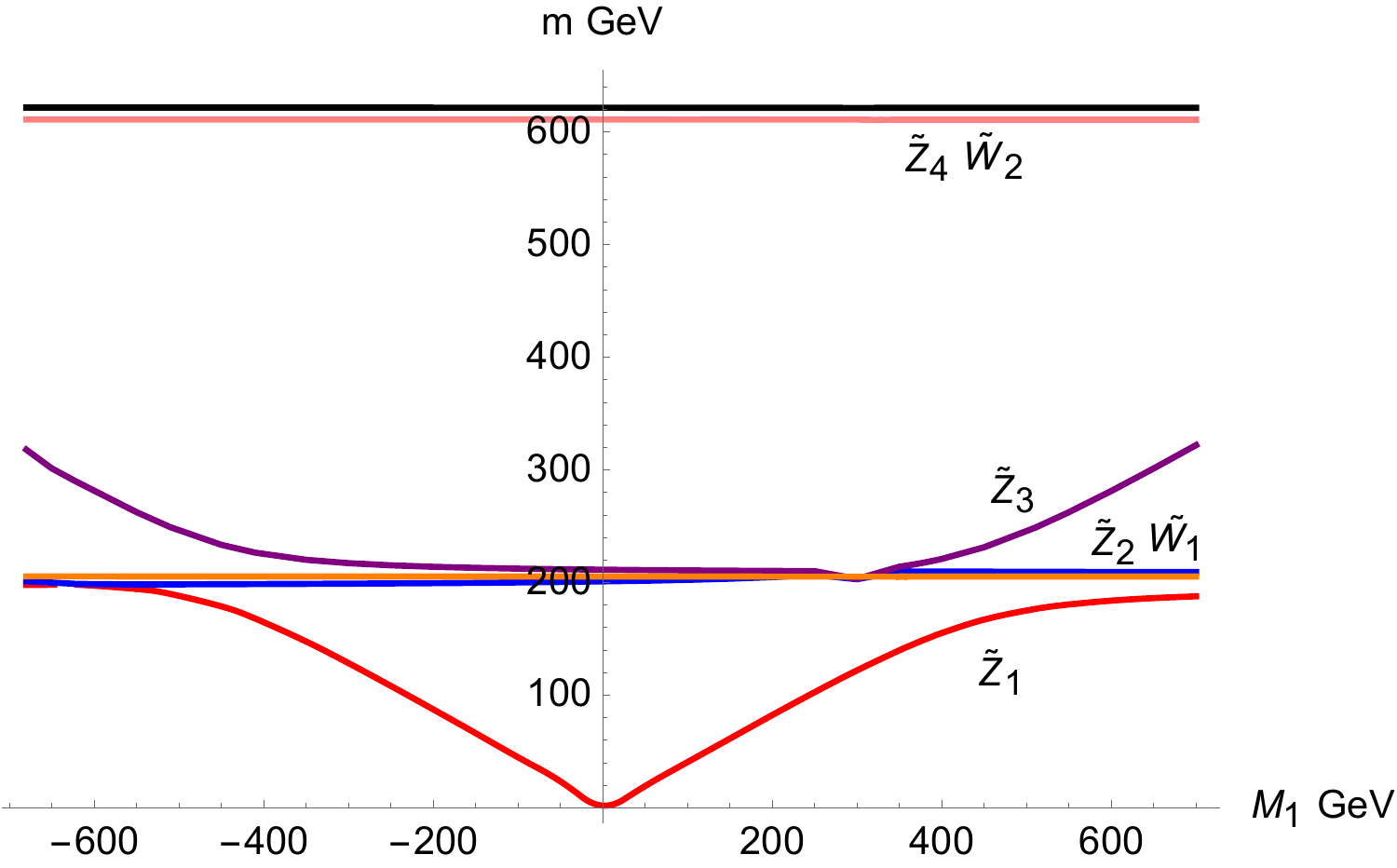}
\end{center}
\caption{Variation of electroweak-ino masses  vs. $M_1$ 
for a general RNS SUSY benchmark model with variable $M_1$ and $M_2=M_3$
\label{fig:mass_M1}}
\end{figure}

In Fig. \ref{fig:mass_M1}, we show the mass values of the charginos and
neutralinos as $M_1$ is varied between -700 GeV to 700~GeV.  For
$M_1=700$~GeV, the gaugino mass unification point, we find that
$\tw_1$ and $\tz_{1,2}$ are all higgsino-like with mass values 
clustered around $\mu= 200$~GeV 
while the bino-like $\tz_3$ lies near 300 GeV and
the wino-like $\tz_4$ and $\tw_2$ lie at $\sim 600$ GeV.  As $M_1$ is
lowered, then the bino component of $\tz_1$ increases while the
bino-component of $\tz_3$ decreases. The mass eigenvalues track the
gaugino/higgsino content, and as we pass through $M_1=300$~GeV, the $\tz_1$
and $\tz_3$ exchange identities and interchange from being bino-like to
higgsino-like. A similar level crossing is seen on the negative $M_1$
side of the figure. Since there is no charged bino, the values of
$m_{\tw_{1,2}}$ remain constant (at $\mu$ and $M_2({\rm weak})$)
with variation of $M_1$. Since the value
of $m_{\tz_1}$ is decreasing as $M_1$ decreases, then the mass gaps
$m_{\tw_1}-m_{\tz_1}$ and $m_{\tz_2}-m_{\tz_1}$ also increase.  The mass
gaps reach values of $\sim 150$ GeV for $M_1$ as small as 50 GeV. This
should render signals from $\tw_1\tz_2$ and $\tw_1\tw_1$ production much
easier to detect at the LHC as compared to the RNSh case. 

\begin{figure}[tbp]
\begin{center}
\includegraphics[height=0.3\textheight]{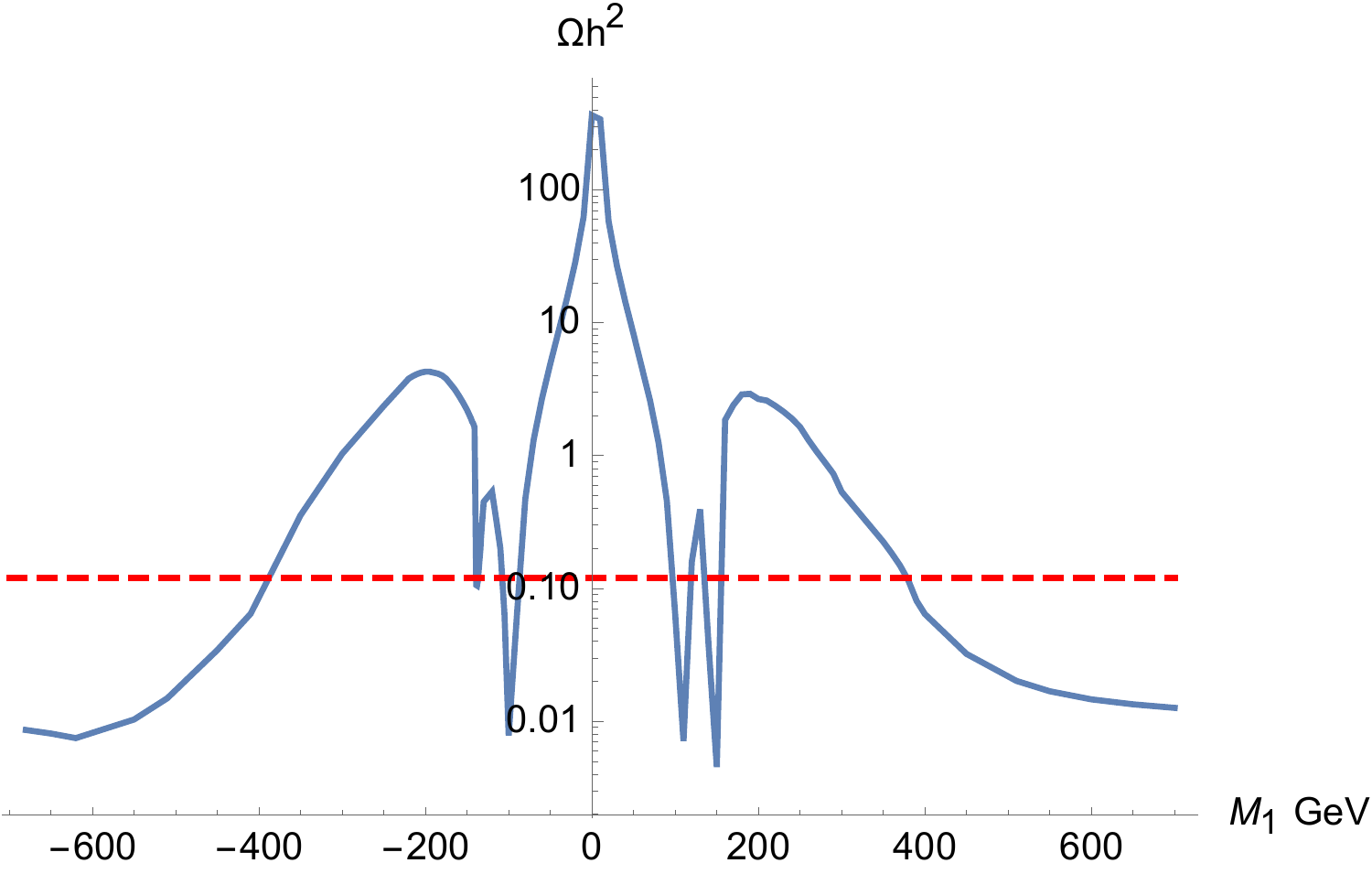}
\end{center}
\caption{Variation of $\Omega_{\tz_1}^{TP}h^2$ vs. $M_1$ 
for a general RNS SUSY benchmark model with variable $M_1$ and
$M_2=M_3$. The dashed line shows the measured value of the cold dark
matter relic density.
\label{fig:oh2_M1}}
\end{figure}

In Fig. \ref{fig:oh2_M1}, we show the thermally-produced neutralino
relic density as calculated using the IsaReD program \cite{isared}. The
value of $\Omega_{\tz_1}h^2$ begins at $\sim 0.01$ for $|M_1|=700$ GeV
which is typical for a higgsino-like LSP of mass 200 GeV.  As $|M_1|$
decreases, then the bino content of $\tz_1$ becomes larger -- 
reducing the annihilation cross section -- 
so that the thermal relic density correspondingly increases.  
For $|M_1|\simeq 380$ GeV, the value of $\Omega_{\tz_1}h^2$ reaches 0.12, {\it i.e.}  it
saturates the measured DM abundance, and we have the so-called
well-tempered neutralino. For even lower values of $|M_1|$, then
neutralinos are unable to annihilate efficiently and $\Omega_{\tz_1}h^2$
exceeds 1 except for special values where the neutralino annihilation
cross-section is resonance-enhanced. For $|M_1|\sim 150$ GeV, then the
bino-like neutralino has mass $m_{\tz_1}\sim m_h/2$ so that neutralinos
can efficiently annihilate through the light Higgs resonance. The
annihilation rate at resonance is not quite symmetric for the two signs
of $M_1$.  For even lower values of $|M_1|$, then $m_{\tz_1}\sim M_Z/2$
so that neutralinos efficiently annihilate through the $Z$ boson
pole. At values of $|M_1|< 100$ GeV we move below the $Z$-resonance
and due to the increasing bino content of $\tz_1$, the LSP annihilation
cross section becomes  even smaller, leading to an even larger thermal relic
density.\footnote{We remind the reader that these parameter
regions with seemingly too large a thermal neutralino relic density
should not summarily be excluded because the neutralino relic density
can be diluted if, for instance, there are heavy particles with late
decays into Standard Model particles in the early universe.}
%
%

We display the SUSY spectrum for $M_1({\rm GUT})=380$ GeV, the value for
which the thermal neutralino relic density $\Omega_{\tz_1}^{TP}h^2$
essentially saturates the measured abundance so that
$\Omega_{\tz_1}h^2=0.12$, in Table \ref{tab:bm} as RNSb.  In this case,
the $\tz_1$ is a bino-higgsino admixture albeit already it is dominantly
bino-like.  The mass gap $m_{\tw_1}-m_{\tz_1}$ is $\sim 56$ GeV while
the mass gap $m_{\tz_2}-m_{\tz_1}$ is $\sim 60$ GeV.

\subsection{Implications for LHC13}

The possibility of non-universal gaugino mass parameters has important
implications for discovery of natural SUSY at LHC13.

\subsubsection{Gluino pair production: multi-jet + $\eslt$ events} 
\label{sec:gl_bino}

Since squarks are very heavy, the multijet + $\eslt$ signal mainly
arises from $pp\to\tg\tg X$ followed by gluino cascade decays mainly via
$\tg\to tb\tw_j$ and $t\bar{t}\tz_i$. For a fixed $m_{\tg}$, but varying
$M_1$, one still expects multi-lepton plus multi-jet$+\eslt$ events at a
rate which mainly depends on the value of $m_{\tg}$.  For discovery via
gluino pair production, the LHC13 reach -- which extends to about
$m_{\tg}\sim 1.7$ TeV (for $m_{\tg}\ll m_{\tq}$) for 100 fb$^{-1}$ of
integrated luminosity \cite{rns@lhc} -- tends to be dominated by
multi-jet$+\eslt$ channel and so changes little compared to the case of
universal gaugino masses.  For the RNS point in question, the gluino
dominantly decays via $\tg \to \tst_1 t$, and the $\tst_1$ subsequently
decays via $\tst_1 \to b\tw_1, t\tz_{1,2,3}$. Within the gluino pair
cascade decay events, the isolated multi-lepton content should increase
with decreasing $M_1$ due to the increased mass gap between
$\tw_1-\tz_1$ and $\tz_{2,3} -\tz_1$ since one may also obtain energetic
leptons from $\tw_1\to \ell\nu_{\ell}\tz_1$ and
$\tz_2\to\tz_1\ell^+\ell^-$ three body decays in addition to those 
from top or $\tz_3$ decays. If $M_1$ is sufficiently
small, then the two-body decays $\tw_1\to\tz_1 W$ and $\tz_2\to\tz_1 Z,\
\tz_1h$ open up.  The latter two decays, if open, tend to occur at
comparable rates in natural SUSY with a bino-like LSP since the lighter
-inos tend to be a gaugino-higgsino admixture.  The isolated
opposite-sign/same flavor (OS/SF) dileptons present in cascade decay
events will have mass edges located at $m_{\tz_2}-m_{\tz_1}$ for
three-body decays, or else real $Z\to \ell^+\ell^-$ 
or $h \to b\bar{b}$
pairs will appear in the case of two-body decays of $\tz_2$ and $\tz_3$:

\subsubsection{Electroweak -ino pair production}

For electroweak-ino pair production, allowing non-universality
in the gaugino sector changes the situation quite
dramatically.  In the case of RNS with gaugino mass unification, the
higgsino pair production reactions $pp\to \tw_1^+\tw_1^-$ and $\tw_1\tz_{1,2}$ 
are largely invisible due to the small mass gaps \cite{rns@lhc}. 
It may, however, be possible to detect higgsino pair production 
making use of initial state QCD radiation and
specially designed analyses if the higgsino mass is below $\sim
170-200$~GeV, depending on the integrated luminosity \cite{kribs,dilep}.

The wino pair production process $pp \to \tw_2\tz_4 X$ 
can lead to a characteristic same-sign
diboson signature \cite{lhcltr} arising from $\tw_2^\mp\to \tz_1W ^\mp$ and
$\tz_4\to\tw_1^\pm W^\mp$ decays, where the higgsinos decay to only soft
visible energy and are largely invisible.

\begin{figure}[tbp]
\includegraphics[height=0.3\textheight]{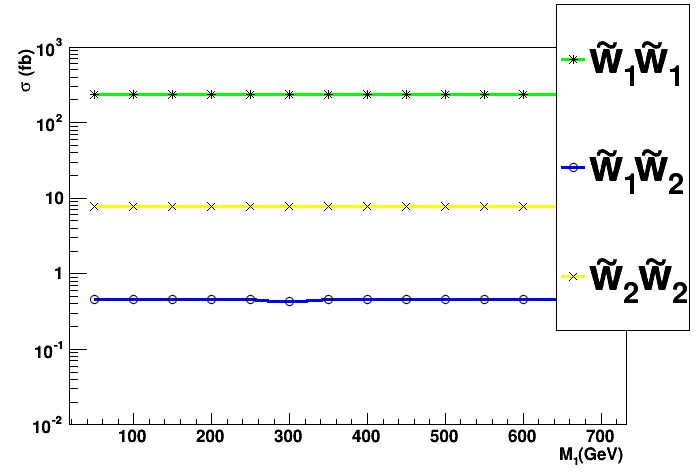}\\
\includegraphics[height=0.3\textheight]{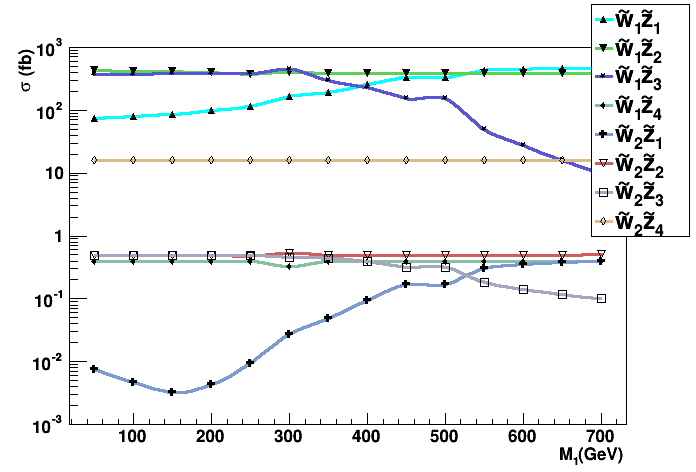}\\
\includegraphics[height=0.3\textheight]{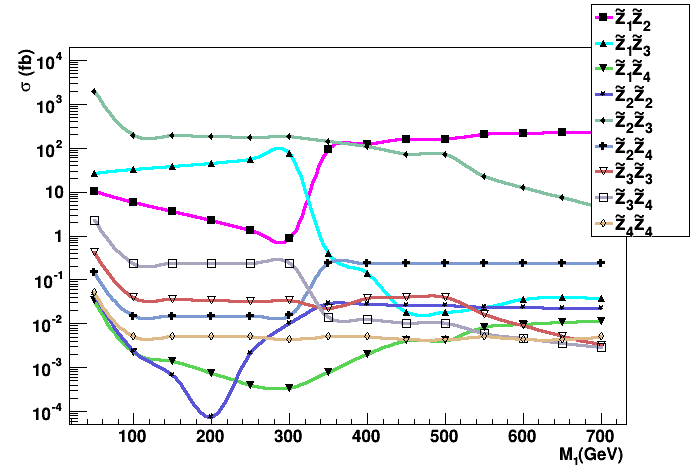}
\caption{Electroweak -ino pair production cross sections versus 
$M_1$ for the RNS SUSY benchmark model with variable $M_1$ but with $M_2=M_3$
\label{fig:sig1}}
\end{figure}

In contrast, as $M_1$ diminishes, then the growing $\tw_1-\tz_1$ and
$\tz_2-\tz_1$ mass gaps give rise increasingly to visible decay products
and a richer set of electroweak -ino signals.  In Fig. \ref{fig:sig1},
we show the NLO cross sections obtained using Prospino \cite{prospino}
for various electroweak-ino pair production reactions versus variable
$M_1(GUT)$ for the RNS benchmark case.\footnote{Since, as we saw in the
previous figures the mixing patterns are roughly symmetric about
$M_1=0$, and because it is relatively time-consuming to run Prospino, we
show results only for positive values of $M_1$.}  As $M_1$ falls to
lower values, the chargino pair rates remain constant since $\mu$ and
$M_2$ do not change.  The $\tw_1\tw_2$ cross section in the topmost
frame is small because squarks are very heavy, and the $Z\tw_1\tw_2$
coupling is dynamically suppressed. Although the $\tw_1\to f\bar{f}'\tz_1$
decay products become more energetic with reducing $|M_1|$, the 
chargino pair signals are
typically challenging to extract from large SM backgrounds such as
$W^+W^-$ production.

For $\tw_1\tz_{1,2}$ production, the cross sections can be large but the
decays give only soft visible energy for $M_1\sim 700$ GeV.  But as
$M_1$ is lowered, the cross section for $\tw_1\tz_2$ remains large but
the mass gaps increase. Ultimately, the clean trilepton 
signature should
become visible against SM backgrounds \cite{trileptev,trileplhc}.  
Also, the reaction $pp\to
\tw_1\tz_3$ has an increasing cross section as $M_1$ decreases and
should give rise to $\ell+Z$ events: trileptons where one pair
reconstructs a real $Z$ \cite{wz}, as is the case for the RNSb benchmark
point: see also Ref's.~\cite{shufang,Martin:2014qra}. 
Ultimately, the $\tz_3\to \tz_1 h$
mode also opens up, reducing the trilepton signal but potentially
offering an opportunity for a search via the $Wh$ channel \cite{wh}.

In models with heavy squarks, higgsino pair production reactions make
the main contribution to neutralino pair production processes. In many
models, $|\mu|$ is large, making neutralino pair production difficult to
see at hadron colliders. Natural SUSY models with non-universal gaugino
masses are an exception as can be seen from the bottom frame of
Fig.~\ref{fig:sig1} where we show cross-sections for various neutralino
pair production processes. The bino-higgsino level crossing that we
mentioned earlier is also evident: for large $M_1$ the $\tz_1$ and
$\tz_2$ are higgsino-like states and $\tz_1\tz_2$ production (solid
squares) dominates, whereas for small $M_1$ then $\tz_2$ and $\tz_3$ are
higgsino-like and $\tz_2\tz_3$ production (left-pointing triangles) is
dominant even though the $\tz_1\tz_2$ and $\tz_1\tz_3$ reactions are
kinematically favoured.  Also $\tz_1\tz_2$ and $\tz_2\tz_3$ production
can lead to dilepton and four-lepton final states which may be visible,
and to $ZZ, Zh$ and $hh +\eslt$ final states if $|M_1|$ is sufficiently
small.

\subsection{Implications for ILC physics}

The prospects for SUSY discovery and precision measurements in the RNS
model have been examined for an International Linear $e^+e^-$ Collider
(ILC) with $\sqrt{s}\sim 250-1000$ GeV in Ref. \cite{ilc}. 
Such a
machine is a higgsino factory in addition to a Higgs factory and
even with small (10 GeV) inter-higgsino mass gaps, SUSY signals should stand
out above SM backgrounds. The clean environment, together with the
availability of polarized electron beams, also allows for precision
measurements that point to the higgsino origin of these events.   The
main reactions of import are $e^+e^-\to \tw_1^+\tw_1^-$ and
$\tz_1\tz_2$ production.

\begin{figure}[tbp]
\begin{center}
\includegraphics[height=0.25\textheight]{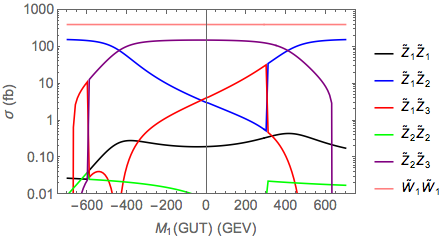}
\end{center}
\caption{Chargino and neutralino production cross sections at a linear 
$e^+e^-$ collider with $\sqrt{s}=500$ GeV with unpolarized beams 
for the RNS SUSY benchmark model with variable $M_1$ but with $M_2=M_3$
\label{fig:e+e-1}}
\end{figure}

In the case where $M_1$ is low enough so that one obtains a bino-like
LSP, the second higgsino state $\tz_3$ also becomes accessible, and
reactions involving $\tz_3$ provide even richer prospects for SUSY
discovery.  Various SUSY pair production cross sections are shown in
Fig. \ref{fig:e+e-1} versus variable $M_1$ and for $\sqrt{s}=500$
GeV. The electron and positron beams are taken  to be
unpolarized in this figure. Once again the level crossings between bino and
higgsino-like states are evident. For the case of unified gaugino masses
with $M_1=700$ GeV, then indeed only $\tw_1\tw_1$ and $\tz_1\tz_2$ are
available. However, as $|M_1|$ is lowered, then $\sigma (\tw_1\tw_1 )$
remains constant although the decay products of $\tw_1$ become more
energetic once the LSP becomes bino-like and lighter than the
higgsino. The dijet mass spectrum from $\tw_1\to \tz_1 q\bar{q}'$ decay
allow for precision extraction of $m_{\tw_1}$ and $m_{\tz_1}$ and also
extraction of the weak scale SUSY parameters $\mu$ and also $M_1$, if
the bino mass is small enough \cite{jlc,bmt,ilc}.

Turning to neutralino production, we see that higgsino pair production
-- $\tz_1\tz_2$ production if $|M_1|$ is large, and $\tz_2\tz_3$
production for small values of $|M_1|$ -- dominates the neutralino cross
section just as in the LHC case.  Notice that for $0< M_1 <
300$~GeV, $\tz_1\tz_3$ production also occurs at an observable rate,
falling with reducing $M_1$ because of the increasing bino content of
$\tz_1$.\footnote{This is somewhat different from the behaviour in
  Fig.~\ref{fig:sig1} where we see, for example, that
  $\sigma(\tz_1\tz_2)$ increases with reducing $M_1$. We attribute this
  to the reduction in mass of the $\tz_1\tz_2$ system and the
  concomitant increase of the parton densities at the LHC.} We have
checked that the strong dip in $\sigma(\tz_1\tz_3)$ around $M_1 \simeq
500$~GeV is due to an accidental cancellation in the $Z\tz_1\tz_3$
coupling.\footnote{The alert reader may wonder why there is no similar
  dip in Fig.~\ref{fig:sig1}. We remark that the code used to make
  Fig.~\ref{fig:e+e-1} uses tree-level masses and mixings among
  charginos and neutralinos, whereas Fig.~\ref{fig:sig1} includes
  effects of radiative corrections to the spectrum. These corrections,
  of course, shift the location as well as depth of the dip. We have
  checked that the coupling is indeed suppressed even with radiative
  corrections, but there is no big dip, at least within resolution of
  the scan. Since it has no implications physics-wise because
  $\sigma(\tz_1\tz_3)$ in Fig.~\ref{fig:sig1} is already very small for
  $M_1 \sim 500$~GeV, we have not attempted to refine this figure.}
$\tz_2\tz_3$ and $\tz_1\tz_3$ production should lead to interesting
event topologies, including $Z+\eslt$ and $h+\eslt$ events where the
missing mass does {\it not} reconstruct to $M_Z$, depending on the decay
modes of the neutralinos. On the negative $M_1$ side, the $\tz_1\tz_3$
cross section is small, except beyond the level crossing at $M_1\simeq
-600$~GeV.
  
Before closing, we note that these neutralino and chargino cross sections
are also sensitive to beam polarization. This can serve to extract
the gaugino/higgsino content of the charginos and neutralinos that are
being produced.

\subsection{Implications for dark matter seaches}

In the RNS model with unified gaugino masses and a higgsino-like LSP,
the relic density of thermally produced neutralinos is much smaller than the
observed density of cold dark matter. This allows for a contribution
from axions \cite{axion} that must  be present if nature adopts the
Peccei-Quinn solution to the strong $CP$ problem. 
In the case of DFSZ axions \cite{dfsz}, one
also gains a solution to the SUSY $\mu$ problem and can allow for a
natural value of $\mu\sim 100-200$ GeV via radiative PQ
breaking \cite{radpq}. In such models, the DM tends to be
axion-dominated \cite{dfsz1} with a local abundance of neutralino WIMPs 
reduced by
factors of 10-15 from usual expectations.  The reduced local abundance
makes direct detection more difficult since detection rates depend
linearly on the local neutralino abundance. Indirect detection rates 
from WIMP halo annihilations depend on the
square of the local abundance so are even more suppressed in models where
the WIMPs only make up a fraction of the dark matter \cite{bbm}.

For the more general model where $|M_1|$ may be lower than expected from
gaugino mass unification, the thermally-produced neutralino abundance is
increased, and consequently one expects a greater fraction of neutralino
dark matter compared to axions, assuming there are no other processes
that affect the neutralino relic density. The increased local neutralino
abundance leads to more favorable prospects for WIMP direct and indirect
detection.

\begin{figure}[tbp]
\begin{center}
\includegraphics[height=0.4\textheight]{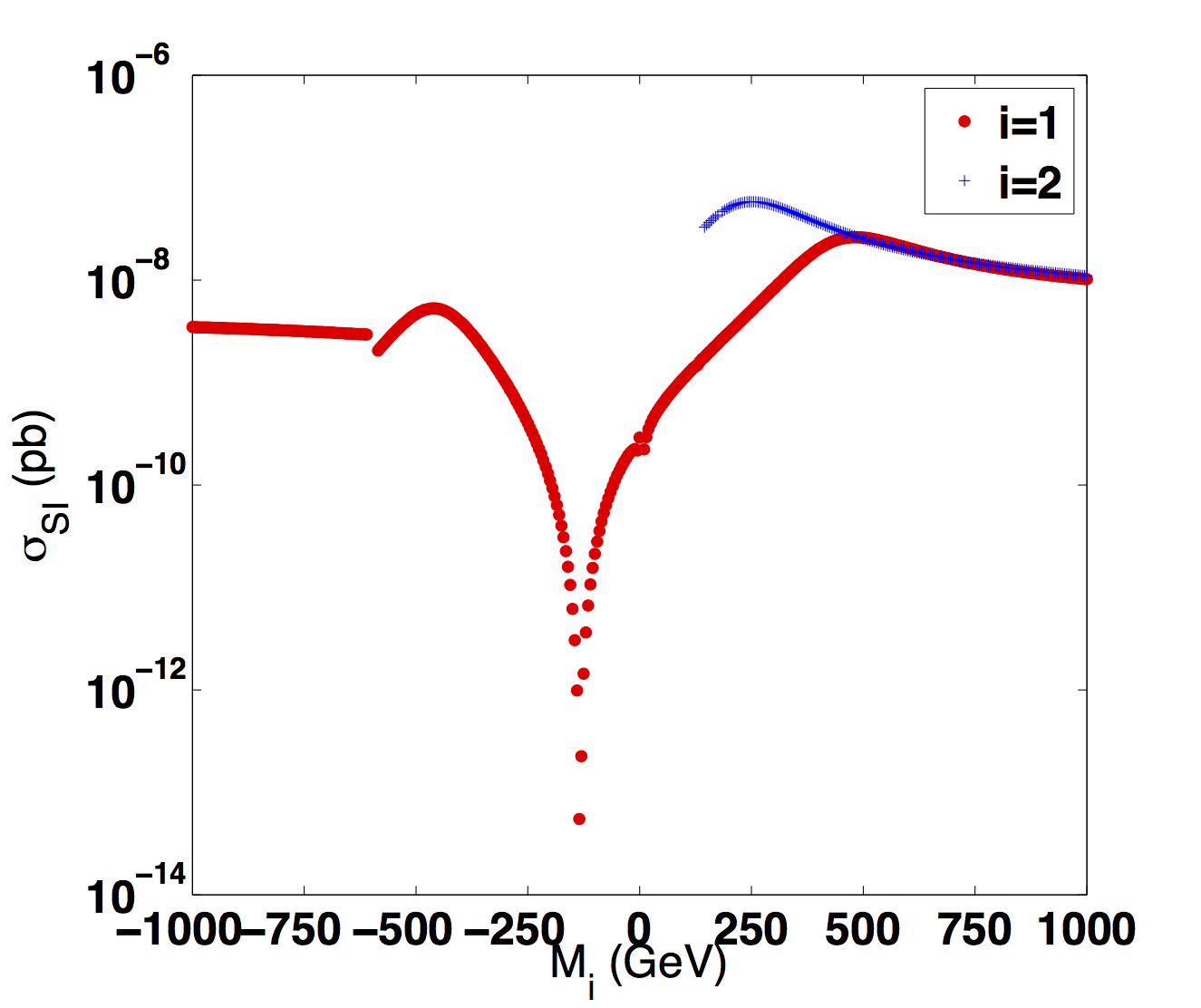}
\end{center}
\caption{Spin-independent $p\tz_1$ scattering cross section
vs. $M_1$ (red dots) or $M_2$ (blue pluses) for the RNS benchmark point.
\label{fig:SI}}
\end{figure}

The spin-independent (SI) WIMP-proton scattering cross section from
IsaReS \cite{isares} is shown in Fig. \ref{fig:SI}. The curve with red
dots shows the case of variable $M_1$.  As $M_1$ decreases from large,
positive values, then the LSP becomes more of a bino-higgsino admixture.
Since the SI cross sections proceeds mainly through light Higgs $h$
exchange, and the Higgs-neutralino coupling is proportional to a product
of gaugino times higgsino components \cite{wss}, then the SI direct
detection cross section increases by up to a factor of $\sim 2$ for
lowered $M_1$. As $M_1$ is lowered even further, then the LSP becomes
more purely bino-like, and the SI direct detection cross section 
drops sharply. 
The sharp dip at $M_1 \simeq -110$~GeV is due to the reduction of the
$h\tz_1\tz_1$ coupling, and also the cancellation between the neutralino
scattering through the exchange of the light CP-even Higgs and that
through the exchange of the heavy CP- even Higgs, denoted as the blind
spot in dark matter direct detection \cite{wss,bsold,bs}.  The kink at
$M_1 \sim -600$~GeV occurs due to a change in the composition of the
LSP: we see from Fig.~\ref{fig:mass_M1} that the levels are getting very
close, and the -inos may be switching composition.

The reader may be concerned that the cross-section in Fig.~\ref{fig:SI}
seemingly violated the upper limits from LUX (Ref. \cite{lux}) of $\sim
(1-2)\times 10^{-9}$~pb for neutralinos in the mass range 20-200~GeV. As
mentioned previously, we should remember that these limits assume that
the LSP saturates the observed density of cold dark matter, which is
certainly {\em not the case} for a higgsino-like LSP (large $|M_1|$
values in the figure). 
Re-scaling the expected event rate by the fractional 
relic density makes the large $|M_1|$ region safe-- though on the edge of
observability-- to LUX constraints (which otherwise assume 
that neutralinos saturate the measured density of cold dark matter).
For smaller values of $|M_1|$, where it may also appear that the direct
detection bound is violated, this clearly is not the case.  We should,
however, keep in mind that for these ranges of $M_1$, the direct
detection rate from which the bound in Ref.\cite{lux} is inferred cannot
be reliably calculated because the physics processes responsible for
bringing the neutralino relic density to its final value lie outside
the present framework. Put differently, we caution against unilaterally
excluding model parameters (including the RNSb model) based on these
considerations, because this frequently requires other assumptions about
the cosmological history of the Universe that have no impact upon
collider physics.\footnote{What is clear from the data is that
neutralinos with a large higgsino content (including the well-tempered
neutralino) cannot be the bulk of the local dark matter.} While WIMP
discovery would be unambiguous, interpretation of the physics
underlying any signal would require a careful specification of all
underlying assumptions. %

\begin{figure}[tbp]
\begin{center}
\includegraphics[height=0.4\textheight]{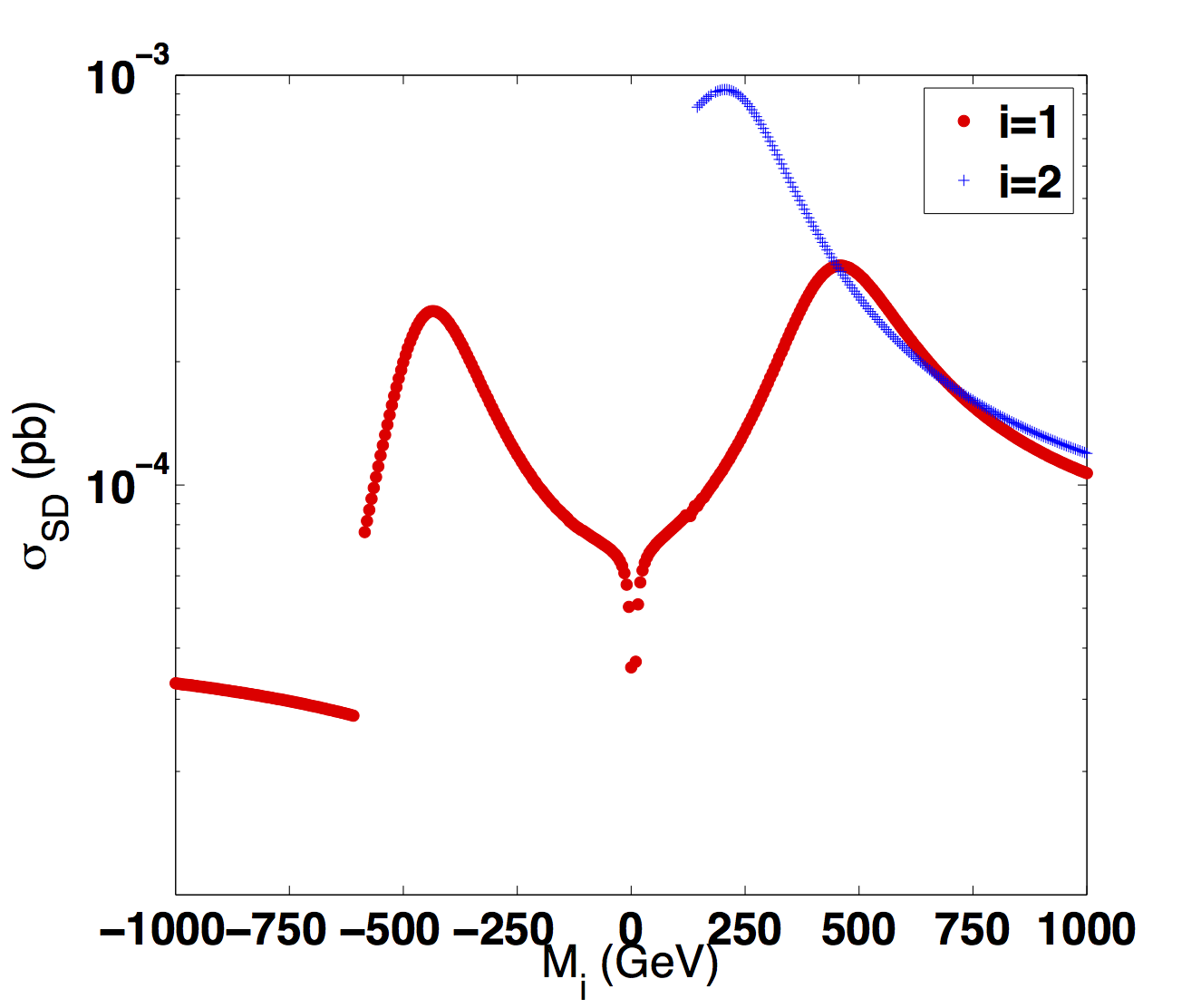}
\end{center}
\caption{Spin-dependent $p\tz_1$ scattering cross section vs. $M_1$ (red
circles) or $M_2$ (blue pluses) for the RNS benchmark point.
\label{fig:SD}}
\end{figure}

The expected spin-dependent (SD) proton-neutralino direct detection
cross section is plotted versus the gaugino mass parameter in
Fig. \ref{fig:SD}. In this case, the scattering occurs dominantly via
$Z$-exchange.  The $Z\tz_1\tz_1$ coupling (Eq. 8.101 of Ref.~\cite{wss}) 
is proportional to
a difference in square of higgsino components of the neutralino. For
$M_1$ large and positive, both higgsino components are comparable and
there is a large cancellation in the coupling. As $M_1$ decreases, the
higgsino components of $\tz_1$ decrease, but the up-type higgsino
content more so than the down type.  There is less cancellation and the
coupling increases. As $M_1$ decreases further, the bino component
increases and the smallness of the higgsino components decreases the
coupling. The negative $M_1$ side shows similar features
until we reach $M_1 \simeq -600$~GeV where the flip in the identity of
the neutralino mentioned in the previous figure results in the discontinuity.

As far as WIMP detection goes, the SD cross section would influence
IceCube \cite{icecube} detection rates the most since the WIMP abundance
in the solar core is determined by equilibration between the capture rate and
the annihilation rate of WIMPs in the sun. 
The scattering/ capture rate of the Sun depends mainly on the
Hydrogen-WIMP scattering cross section which proceeds more through the
SD interaction since there is no nuclear mass enhancement. While some of
the predicted values (red points) might well be marginally excluded by
the IceCube search, the take-away message is that for the most part the
model with $\mu=200$~GeV is on the edge of detectability, as long as
neutralinos dominantly annihilate to $W$ pairs and assuming
that neutralinos essentially saturate the entire cold dark matter relic density.

\begin{figure}[tbp]
\begin{center}
\includegraphics[height=0.4\textheight]{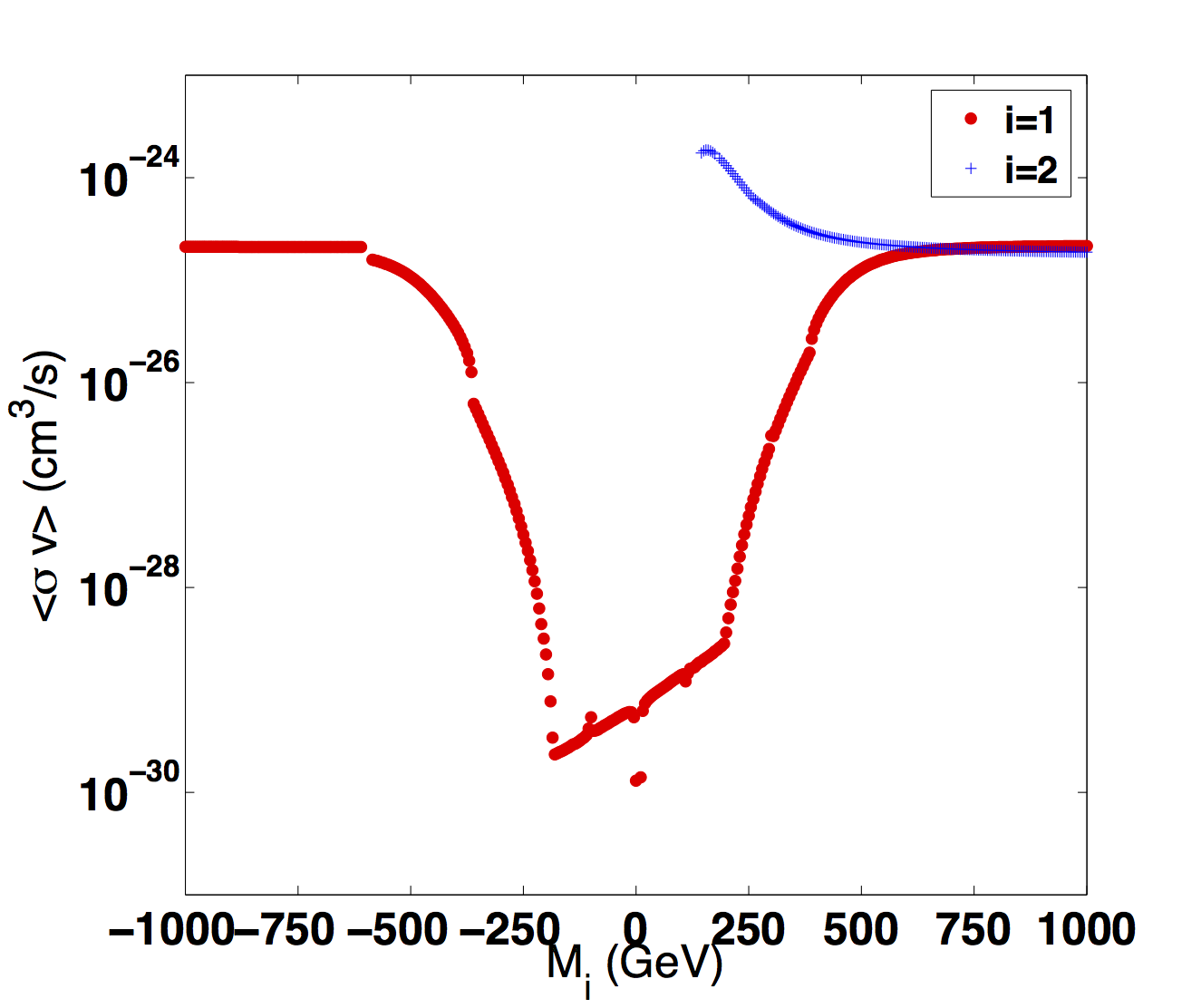}
\end{center}
\caption{Thermally-averaged neutralino annihilation cross section times
velocity at $v=0$ vs. $M_1$ (red dots ) or $M_2$ (blue pluses) for the
RNS benchmark point.
\label{fig:sv}}
\end{figure}

In Fig. \ref{fig:sv}, we show the thermally-averaged neutralino
annihilation cross section times relative velocity evaluated as $v\to
0$. This quantity enters the halo WIMP annihilation rate, and
detection rate for galactic positrons, anti-protons and gamma rays from
WIMP halo annihilations are proportional to this factor.  In the case of
gaugino mass unification where we have a higgsino-like neutralino, then
the local abundance is reduced and the expected detection rate is
reduced by the square of the WIMP underabundance: $\xi^2$ where
$\xi=\Omega_{\tz_1}h^2/0.12$. From the figure, we see that while the
local abundance increases as $|M_1|$ is reduced (Fig. \ref{fig:oh2_M1}), 
the annihilation rate decreases because
annihilation to $WW$s occurs mainly via the (reducing) higgsino
component of the LSP. Once this channel is closed (around $|M_1|\simeq
200$~GeV), annihilation to fermions takes over and the rate drops
further. The FERMI-LAT collaboration has obtained upper limits 
located at 
about a few $\times10^{-26}cm^3/s$ ($\sim 2\times
10^{-25}~cm^3/s$) for annihilation to $b\bar{b}$ ($WW$
pairs) \cite{fermi}. Assuming a Navarro-Frenk-White profile for dwarf
galaxies in the analysis, models with a larger cross section would have
led to a flux of gamma rays not detected by the experiment. Even without
the $\xi^2$ scaling noted above, and certainly after the scaling, these
bounds do not exclude any of the points in the figure.  For completeness
we note that all the caveats that we discussed for the applicability 
of direct detection
bounds are also applicable in this case, and we urge the reader to
use caution in excluding ranges of parameters even if the Fermi
Collaboration obtains tighter bounds in the future.
\section{Natural SUSY with a wino-like LSP}
\label{sec:wino}

In this section, we examine the phenomenological implications of
altering the $SU(2)$ gaugino mass parameter $M_2$ while keeping
$M_1=M_3=700$ GeV.  We begin by showing, as blue pluses, the variation
of $\Delta_{\rm EW}$ with $M_2$ in Fig. \ref{fig:M12_EW}. Again, we see
that $\Delta_{\rm EW}$ is relatively insensitive to $M_2$ except for the
largest values of this parameter.
This is due to the increasing contribution of winos to
$\Sigma_u^u(\tw_{1,2})$.  Thus, models with $M_2\ll M_{1,3}$ lead to a
wino-like LSP at little cost to naturalness.  
For  $M_2< 150$~GeV the chargino becomes lighter than 100~GeV (roughly the
chargino mass bound from LEP2).  Here, and in subsequent figures, we do not
consider negative values of $M_2$ as these lead to a chargino LSP: 
$m_{\tw_1}<m_{\tz_1}$.

\begin{figure}[tbp]
\begin{center}
\includegraphics[height=0.3\textheight]{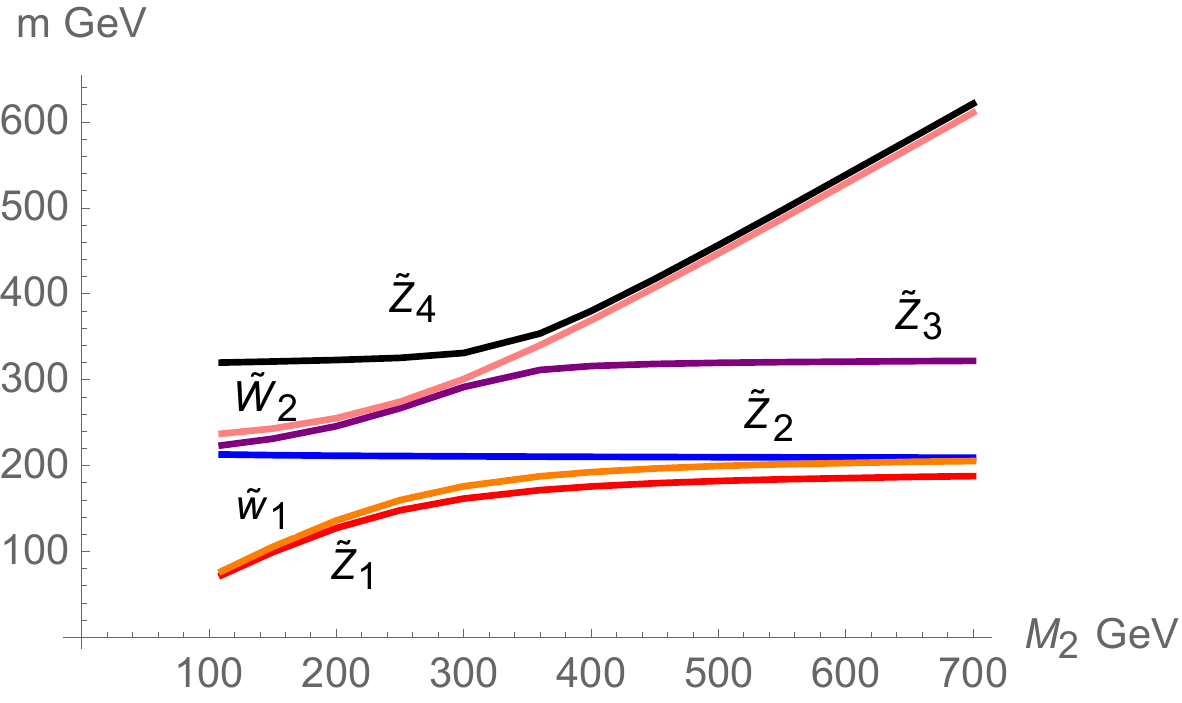}
\end{center}
\caption{Variation of chargino and neutralino masses vs. $M_2$ 
for the RNS SUSY benchmark model with variable $M_2$ but with $M_1=M_3$
\label{fig:mass_M2}}
\end{figure}

In Fig. \ref{fig:mass_M2}, we show how the masses of charginos and
neutralinos change as $M_2$ is reduced from its unified value.  Starting
with the RNSh spectra at $M_2=700$ GeV, where the $\tw_2$ and $\tz_4$
are essentially winos, and $\tz_1$, $\tz_2$ and $\tw_1$ are higgsinos,
we see that as $M_2$ is lowered, the mass of the wino-like states
reduces whereas the higgsino-like states remain with the mass fixed
close to $\mu$. The mass of the bino-like $\tz_3$ also remains nearly
constant. This behaviour persists until we reach the bino-wino level
crossing near $M_2 \simeq 350$~GeV where $\tz_3$ and $\tz_4$ switch
identities. For still lower values of $M_2$, we see another level
crossing between the charged as well as neutral wino-like and higgsino-like
states. For $M_2 < 200$~GeV, the lighter chargino as well as the LSP are
wino-like, the heavier chargino and the neutralinos $\tz_{2,3}$ are
higgsino-like, and $\tz_4$ is mainly a bino. 
The mass gap $m_{\tw_1}-m_{\tz_1}$
has actually decreased with decreasing $M_2$ since these wino-like
states have very tiny mass splittings. The mass gaps $m_{\tw_2}-m_{\tz_1}$ and
$m_{\tz_2}-m_{\tz_1}$ greatly {\it increase} with decreasing $M_2$,
reflecting the widening higgsino-wino mass difference. This should
make their visible decay products harder so that these states 
are  easier to detect at the LHC.

\begin{figure}[tbp]
\begin{center}
\includegraphics[height=0.3\textheight]{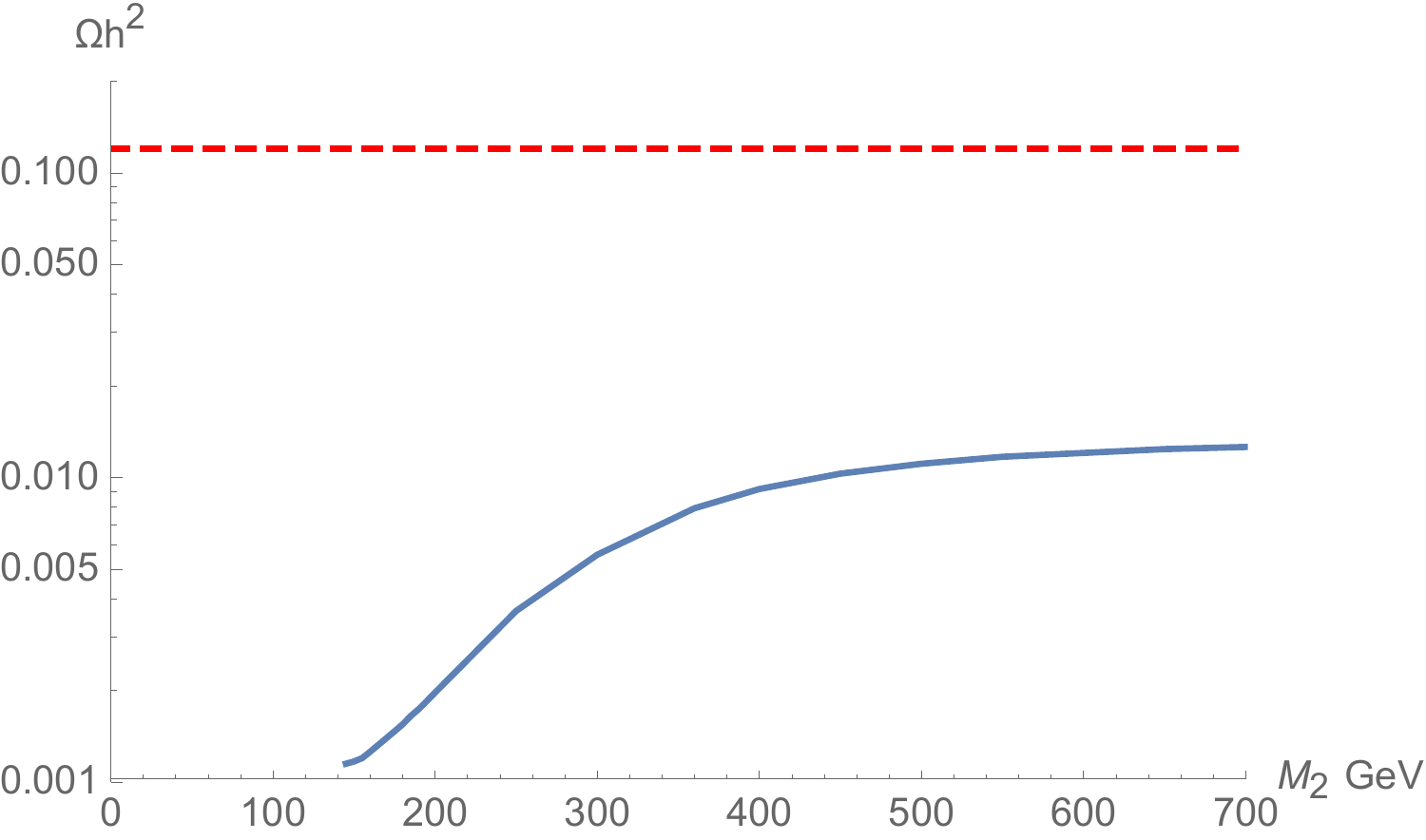}
\end{center}
\caption{Variation of $\Omega_{\tz_1}^{TP}h^2$ vs. $M_2$ (blue curve)
for the RNS SUSY benchmark model with variable $M_2$ but with
$M_1=M_3$. We cut the graph off at the low end because $m_{\tw_1}$ falls
below its LEP2 bound. 
\label{fig:M2_Oh2}}
\end{figure}

We show the thermally-produced neutralino relic density
$\Omega_{\tz_1}h^2$ versus $M_2$ in Fig. \ref{fig:M2_Oh2}. Starting with
$M_2=700$ GeV for which $\Omega_{\tz_1}h^2\sim 0.01$, we see that
$\Omega_{\tz_1}h^2$ steadily decreases with decreasing $M_2$ and reaches
a value $\Omega_{\tz_1}h^2\sim 0.001$ for very low values of $M_2$ where
the $\tz_1$ is nearly pure wino. This is because wino annihilation
proceeds via the larger SU(2) triplet coupling to electroweak
gauge  bosons while
annihilation of higgsinos proceeds via the smaller doublet coupling --
 the cross section for annihilation to $W$ pairs, which is
dominated by the $t$-channel chargino exchange, goes as the fourth power
of this coupling. Thus, in the case of low
$M_2$ with a wino-like neutralino, we might expect an even more reduced
local abundance from {\em thermally produced LSPs}. The balance may be 
made up either by axions or other relics, or by LSPs produced 
by late decays of heavier particles. 
We cut the graph off when $m_{\tw_1}$ falls below its LEP2 bound. 
We do not see any dips
corresponding to $s$-channel 
$h$ or $Z$ funnel annihilation as these fall in the LEP2 excluded
region.  
%

An RNS benchmark point with a wino-like LSP is shown in Table
\ref{tab:bm} and is labelled as RNSw. All input parameters for RNSw are
the same as for RNSh except now $M_2$ is chosen to be 175 GeV. The
$\tw_1-\tz_1$ mass gap has decreased to just 7.3 GeV while the
$\tz_2-\tz_1$ mass gap has increased beyond the RNSh value up to $\sim
97$ GeV, large enough so that both $\tz_2\to\tz_1 Z$ and $\tz_2 \to
\tw_1^\pm W^\mp$ decays are now allowed. In such a scenario, we would
expect LHC SUSY cascade decay events to be rich in content of real $Z$
bosons that could be searched for at the LHC. In fact, the CMS
\cite{cms_ino} and ATLAS \cite{atlas_ino} collaborations have already
obtained bounds on chargino and neutralino masses from an analysis of
about 20~fb$^{-1}$ of LHC8 data. These limits are obtained in simplified
models from an analysis of expectations from $\tw_1\tz_2$ and
$\tw_1\tw_1$ production at LHC8, assuming that $m_{\tw_1}=m_{\tz_2}$ and
that the charginos (neutralinos) decay 100\% of the time to $W$ bosons
($Z$ bosons or Higgs bosons). The ATLAS bound\cite{atlas_ino} -- obtained
from a combination of the dilepton and trilepton channels -- excludes wino
pair production for wino masses up to 250 (400)~GeV provided the LSP is
lighter than 100 (150)~GeV, while the current CMS limit is
considerably less restrictive.  While these limits are not directly
applicable to pair produced $\tw_1$ and $\tz_2$ for the RNSw scenario in
the table, the reader may be concerned that {\em higgsino-pair
production} processes $pp \to \tw_2\tz_{2,3}X, \tw_2\tw_2X$ would lead
to final states similar to those that the LHC searches look for.  
It is clear that the RNSw scenario, with $m_{\tz_1}=114$~GeV, is
clearly allowed by current searches: aside from the fact that the LSP
mass exceeds 100 GeV for which there is no LHC limit,  the
higgsino pair production cross section is smaller than that for wino
pair production. This will further weaken the bound for the RNSw case.  Data
from the LHC13 run should, however, decisively probe this benchmark
point.

\subsection{Implications for LHC13}

\subsubsection{Gluino pair production: multijet plus $\eslt$ events}

As discussed in Sec.~\ref{sec:gl_bino}, the discovery reach of LHC13 for
gluino pairs mainly depends on the value of $m_{\tg}$ which dictates the
total $\tg\tg$ production cross section in the case of heavy squarks. We
would thus expect a similar LHC13 reach for gluino pair production in
the RNSw case as for RNSh and as for mSUGRA/CMSSM for comparable gluino
masses and heavy squarks. Also, in the RNSw case, then charginos $\tw_1$
will still be largely invisible due to their soft decay products. In
some AMSB models with a wino-like LSP, then the mass gap
$m_{\tw_1}-m_{\tz_1}$ lies at the 100 MeV level leading to long-lived
winos whose tracks before decay may be visible \cite{amsbpheno}. In our
case though , since $\mu$ is 100-200~GeV as required by naturalness, the
$\tw_1-\tz_1$ mass gap tends to lie in the 5-10 GeV range and so charged
winos will be short-lived with no discernable tracks or kinks.  However,
in the RNSw case, then the $\tz_2-\tz_1$ mass gap does become large and
the well-known dilepton mass edge at $m_{\tz_2}-m_{\tz_1}$ should be
observable for energetic enough $\tz_2\to\tz_1\ell^+\ell^-$ decays if
$m_{\tz_2}-m_{\tz_1} < M_Z$. In the case where the decay $\tz_2\to\tz_1
Z$ opens up, then the gluino cascade decay events (which, depending on
the spectrum, should mostly proceed via real or virtual stop decays
because stops are much lighter than first/second generation squarks)
should be rich in OS/SF dileptons which reconstruct $M_Z$. Note also
that for modest values of $M_2$, then $\tz_3$ is also expected to be
relatively light, and should also be accessible via gluino decays.  For
yet smaller values of $M_2$, $\tz_{2,3}\to\tz_1 h$ may also be allowed
and should occur with a comparable branching fraction to the decay to
real $Z$s.

\subsubsection{Electroweak-inos at LHC13}

\begin{figure}[tbp]
\begin{center}
\includegraphics[height=0.3\textheight]{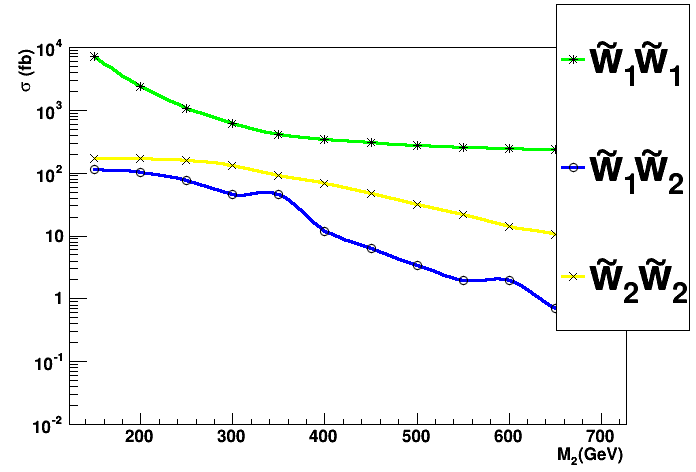}\\
\includegraphics[height=0.3\textheight]{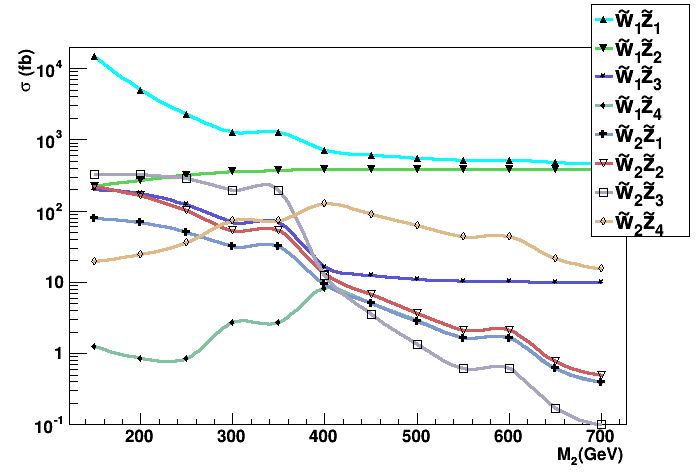}\\
\includegraphics[height=0.3\textheight]{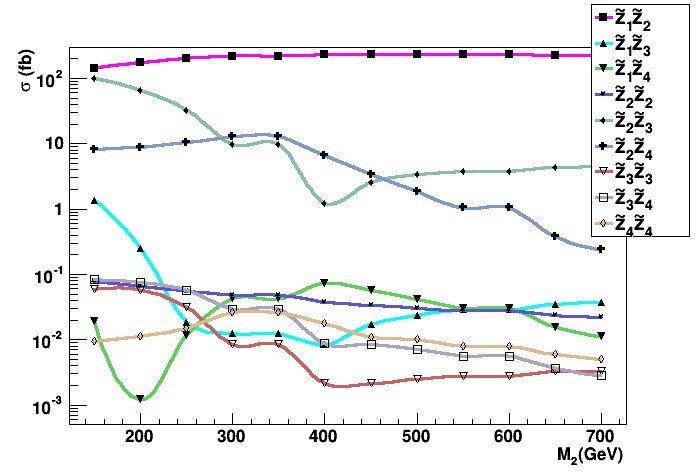}
\end{center}
\caption{Electroweak-ino pair production cross sections versus 
$M_2$ for the RNS SUSY benchmark model with variable $M_2$ but with $M_1=M_3$
\label{fig:xs_m2}}
\end{figure}

In Fig. \ref{fig:xs_m2}, we show NLO cross sections from
Prospino \cite{prospino} for electroweak -ino pair production at LHC13
for the RNS benchmark but for variable $M_2$.  Chargino pair production --
shown in the topmost  frame -- occurs via wino as well as via higgsino pair 
production. For large $M_2$ the latter dominates, but as $M_2$ is
reduced, wino pair production increases in importance until it
completely dominates
for $M_2 \sim 100$~GeV. $\tw_1\tw_2$ production, for the most part
occurs via small gaugino/higgsino content, and so has a smaller cross
section than the kinematically disfavoured $\tw_2\tw_2$ production. 
The level crossing as the light chargino transitions from being
higgsino-like to wino-like as $M_2$ reduces is also evident in the upper
two curves.

Chargino-neutralino production, shown in the middle frame, also occurs
via wino as well as higgsino-pair production processes. For large values
of $M_2$, higgsino pair production dominates and $\tw_1\tz_{1,2}$
production processes have the largest cross sections. For very small
values of $M_2$, pair production of winos is dynamically and
kinematically favoured, and $\tw_1\tz_1$ occurs at the highest rate.
The higgsino-like states $\tw_2, \tz_{2,3}$ have masses $\mu$ and are
also produced with substantial cross sections. Notice that $\tw_1\tz_2$
production remains significant even for small values of $M_2$,
presumably because it is favoured by kinematics (and increased parton
luminosity).

Neutralino pair production (shown in the bottom frame) can only occur via
higgsino pair production since electroweak gauge invariance precludes a
coupling of $Z$ to neutral gauginos. As a result, $\tz_1\tz_2$
production dominates for large $M_2$. For small values of $M_2$ (where
$\tz_1$ becomes wino-like) $\tz_2\tz_3$ production becomes important;
however, $\tz_1\tz_2$ production remains large because of large parton
densities.

We see that for $M_2 \alt 300$~GeV, the cross sections 
for $\tw_1\tz_1$ and $\tw_1\tw_1$
production processes increase rapidly with decreasing $M_2$
since $\tw_1$ and $\tz_1$ become increasingly wino-like.  
%
However, since the $\tw_1-\tz_1$ mass gap reduces even below the
higgsino-LSP case, these states remain difficult -- perhaps impossible -- 
to detect.  Possibly $\tw_1\tw_1$ production may be detectable via
vector-boson fusion-like cuts in events where energetic jets with a
large rapidity gap are required \cite{bhaskar}.  Although the cross
section for wino-like $\tw_1\tz_1$ production becomes very large at low
$M_2$, this process is difficult to detect. However, $\tw_1\tz_2$
production remains at viable rates even for low $M_2$.  In this case,
one might look for relatively hard OS/SF dileptons from $\tz_2$ decay
recoiling against only soft tracks and $\eslt$. Other possibly more
promising reactions at low $M_2$ include $\tw_2\tz_3$, $\tw_2\tz_2$,
$\tz_2\tz_{3}$ and maybe also $\tz_2\tz_4$ production, since the decay
products from both the chargino and neutralino should be relatively hard
and can lead to $\eslt$ events with three or more leptons, or
real Z and Higgs bosons. 
 As we mentioned,  LHC collaborations are already searching
for an excess of just such events \cite{cms_ino,atlas_ino,atlas_hino}.
Constraints from $Wh+\eslt$  analyses are currently much weaker
than those from the $WZ+\eslt$ analyses discussed above.
Note also that $\tw_1\tz_3$ and $\tz_1\tz_2$ production each
has a cross section in excess of 100~fb at low $M_2$ but would be
considerably more difficult to detect.
%

\subsection{Implications for ILC}

\begin{figure}[tbp]
\begin{center}
\includegraphics[height=0.3\textheight]{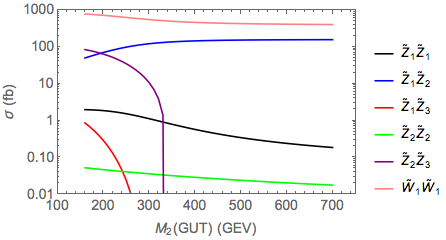}
\end{center}
\caption{Chargino and neutralino production cross sections 
with unpolarized electron and positron beams at a linear 
$e^+e^-$ collider with $\sqrt{s}=500$ GeV 
for the RNS SUSY benchmark model with variable $M_2$ but with $M_1=M_3$
\label{fig:e+e-2}}
\end{figure}

At ILC, the natural SUSY scenario with low $M_2$ becomes both more
challenging and richer.  The cross sections for chargino and neutralino
pair production at ILC500 are shown in Fig. \ref{fig:e+e-2} for
unpolarized beams.  For $M_2=700$ GeV, we have the higgsino pair
production reactions $e^+e^-\to \tw_1^+\tw_1^-$ and $\tz_1\tz_2$
dominating. As $M_2$ is lowered, then the $\tw_1$ becomes more wino-like
and lighter leading to a larger cross section. However, the mass gap
$\tw_1-\tz_1$ drops below 10 GeV making chargino pairs more difficult
but likely still possible to detect with specially designed cuts. Beam
polarization would serve to ascertain the higgsino/wino content of the
chargino.  Also, the $\tz_1\tz_2$ reaction falls with decreasing $M_2$
as the $Z-\tz_1-\tz_2$ coupling decreases ($Z$ only couples to higgsino
components).  As $M_2$ falls below 300 GeV, the the $\tz_2\tz_3$
reaction turns on and grows in importance because the $\tz_3$ becomes
increasingly higgsino-like. Here, we expect $\tz_3$ to decay via 2-body
modes into Z-bosons or higgs bosons and $\tz_2$ to decay either to 2- or 3-body
modes depending on the mass gap. This reaction should be distinctive and
easily visible.

\subsection{Implications for WIMP detection}

WIMP detection for models with radiatively-driven naturalness and a
wino-like WIMP may be either more and less difficult than the case with
gaugino mass unification since, though the nucleon neutralino
scattering cross section is larger,  the local abundance for a thermally
produced wino-like LSP is below the already low value 
typical of a higgsino-like LSP.  Of course, the thermal wino
abundance can be augmented by non-thermal processes involving 
moduli decay \cite{rm} or axino/saxion decay \cite{review} in the early universe.

In Fig. \ref{fig:SI} we show the SI direct detection $\tz_1p$ scattering
cross section versus $M_2$ as the curve with blue pluses. Starting off
at large $M_2$, we see that as $M_2$ is decreased, the
$\sigma^{SI}(\tz_1 p)$ cross section increases, and the increase is
substantially larger than the case of a bino-like LSP. Recall this cross
section proceeds mainly via light $h$ exchange which depends on a
product of gaugino and higgsino components of the neutralino LSP
\cite{wss}. In this case, the wino-component, which involves the larger
$SU(2)$ gauge coupling $g$, becomes enhanced leading to the large cross
section. For small enough $M_2<250$ GeV, the cross section turns around
and decreases with decreasing $M_2$ since the $\tz_1$ becomes more
purely wino-like and the higgsino components are diminished. We note
here that though the cross section in Fig.~\ref{fig:SI} exceeds the
stated bounds ($1-2\times 10^{-9}$~pb for $m_{\tz_1} =100-200$~GeV) in
Ref.~\cite{lux}, these bounds are not directly applicable because they
were obtained assuming the neutralino constitutes the entire dark matter
content of the Universe. For the natural SUSY scenario, the rates in
direct detection experiments could be much smaller, as these scale by
the neutralino fraction of the total local dark matter density. A
wino-like neutralino that forms the bulk of the local dark matter would
be excluded.

In Fig. \ref{fig:SD}, we show the spin-dependent direct detection cross
section $\sigma^{SD}(\tz_1 p)$ versus $M_2$ as the blue curve.  Here,
the SD scattering cross section which proceeds mainly by $Z$ exchange
becomes large since there is less cancellation in the
$Z-higgsino-higgsino$ coupling. For small enough $M_2$, then again the
cross section turns over and decreases due to the diminishing higgsino
components. We see that the cross section exceeds its 90\% CL IceCube
upper limit $\sim 1.5\times 10^{-4}$~pb \cite{icecube} obtained assuming
that LSPs in the sun annihilate dominantly to $W$-pairs if $M_2<
700$~GeV. As discussed earlier, the expected event rate
must be re-scaled by $\xi$ ($= 0.01-0.1$ for thermally produced wino LSPs),
before comparing with IceCube limits.
Then the IceCube limit on the cross section will be correspondingly
degraded, assuming that the neutralino density in the sun is determined
by equilibrium between capture and annihilation rates. The RNSw scenario
satisfies the IceCube bound assuming that the wino relic density is
close to its thermally produced value and that the axion or some other
particle makes up the remainder of the dark matter. Models where the
dark matter is dominantly a wino-like neutralino are strongly excluded
by IceCube.
%

In Fig. \ref{fig:sv} we show $\langle\sigma v\rangle |_{v\to 0}$ versus
$M_2$ as the blue shaded curve. In this case, as $M_2$ falls, then
$\tz_1\tz_1\to WW$ becomes large and the annihilation rate
increases. One might expect increased liklihood for 
indirect WIMP detection via gamma rays and antimatter detection. 
However, the increased annihilation rate is
counter-balanced by a likely decreasing local WIMP abundance where the
detection rate is proportional to the square of the reduced local
abundance. We see that although the predicted rate naively exceeds the upper
limit from Fermi-LAT in Ref.~\cite{fermi}, 
after the $\xi^2$ scaling discussed above exclusion is not possible.

\section{General results in $M_1$ vs. $M_2$ plane}
\label{sec:gen}

While it is instructive to examine natural SUSY models with reduced GUT
scale bino- or wino- mass parameters, there is no compelling reason to
believe that one parameter is unified with $M_3({\rm GUT})$ while the
other is quite different. 
In general one may have arbitrary gaugino masses and
in fact both may be reduced leading to a mixed bino-wino-higgsino LSP. 
Here, we present some illustrative studies of this more general
situation. We can choose $M_1>0$ by convention. The signs of $M_2$ and
$M_3$ as well as $\mu$ are then physically relevant. Since our purpose
is to give a broad brush idea of how RNS phenomenology of
electroweak-inos may be altered, we will take $M_3$ and $\mu$ to be
fixed at their values for the RNSh benchmark point and display results
in the $M_1-M_2$ plane.\footnote{The electroweak sector should be almost
insensitive to $M_3$, but will, of course, be sensitive to the sign of $\mu$.}

\begin{figure}[tbp]
\begin{center}
\includegraphics[height=0.3\textheight]{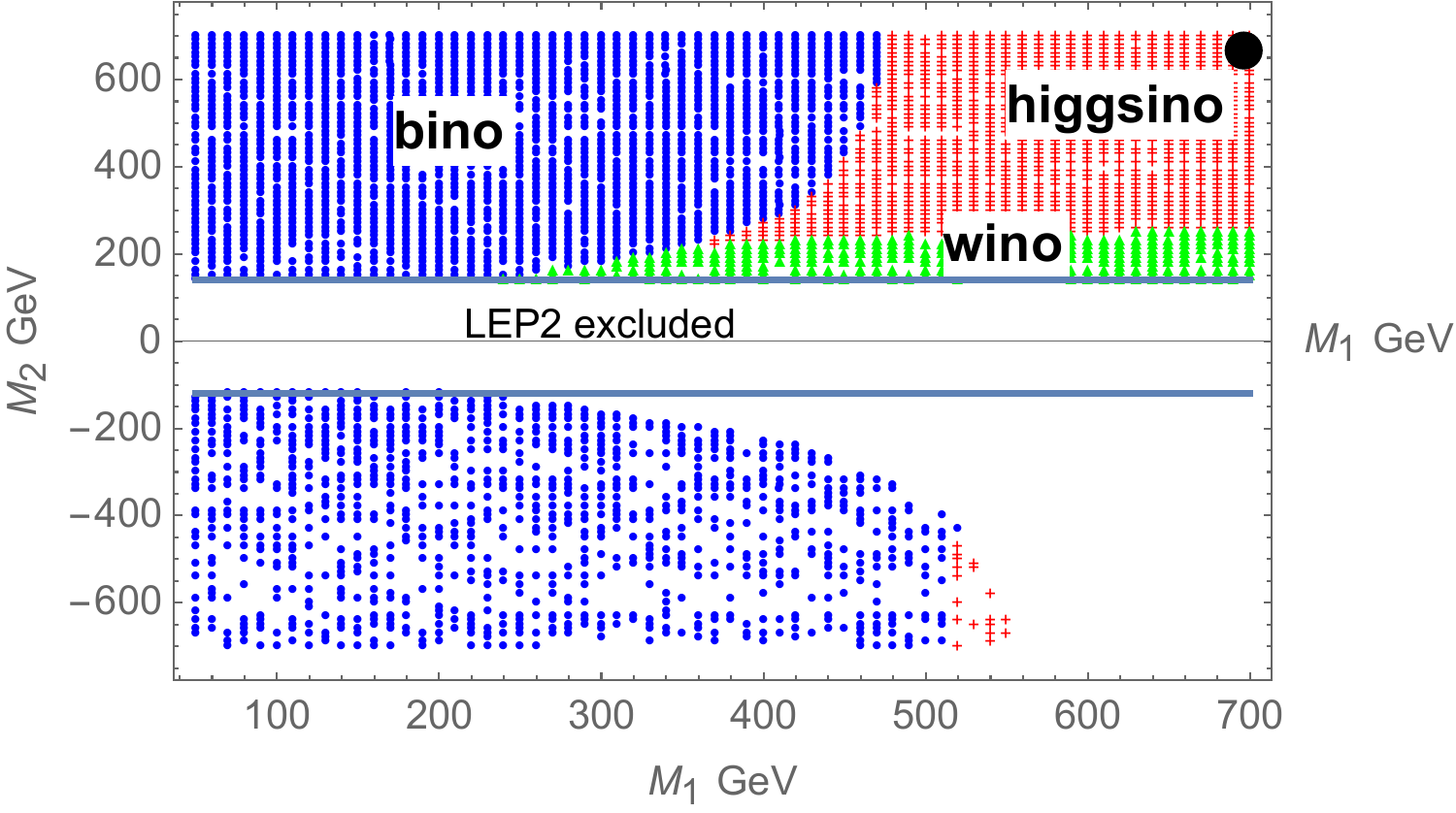}
\end{center}
\caption{Dominant component of the neutralino LSP in the $M_1$ vs. $M_2$
plane for the RNS SUSY benchmark model. The LSP is dominantly a 
bino, wino or higgsino in the region denoted by blue
dots, red pluses  and
green crosses, respectively. Other parameters are fixed at their values
for the RNSh model point in Table~\ref{tab:bm}. In the region
marked LEP2 excluded,
$m_{\tw_1}< 100$~GeV, whereas in the
remaining unshaded region of the lower half plane, $m_{\tw_1} < m_{\tz_1}$. 
\label{fig:LSP}}
\end{figure}

In Fig. \ref{fig:LSP}, we show the $M_1$ vs. $M_2$ plane for the RNS
benchmark model but with $M_1$ and $M_2$ as free parameters. The black
dot in the upper-right corner denotes the location for unified gaugino
masses. The regions of the plot are coded according to the dominant
content of the $\tz_1$: bino (blue dots), wino (green triangles) and higgsino
(red pluses).  The special cases of the previous sections correspond
to moving horizontally to the left or vertically down from the unified
gaugino mass point.  We start the scans at $M_1=50$~GeV and scan both
signs of $M_2$. Here, and in subsequent figures, the band with $|M_2| \alt
150$~GeV is excluded by the LEP2 bound on the chargino. 
In the half plane with $M_2< 0$, the additional region without any 
shading corresponds to a charged LSP ($m_{\tw_1}<m_{\tz_1}$) and so is
excluded by cosmological considerations. 

\begin{figure}[tbp]
\begin{center}
\includegraphics[height=0.4\textheight]{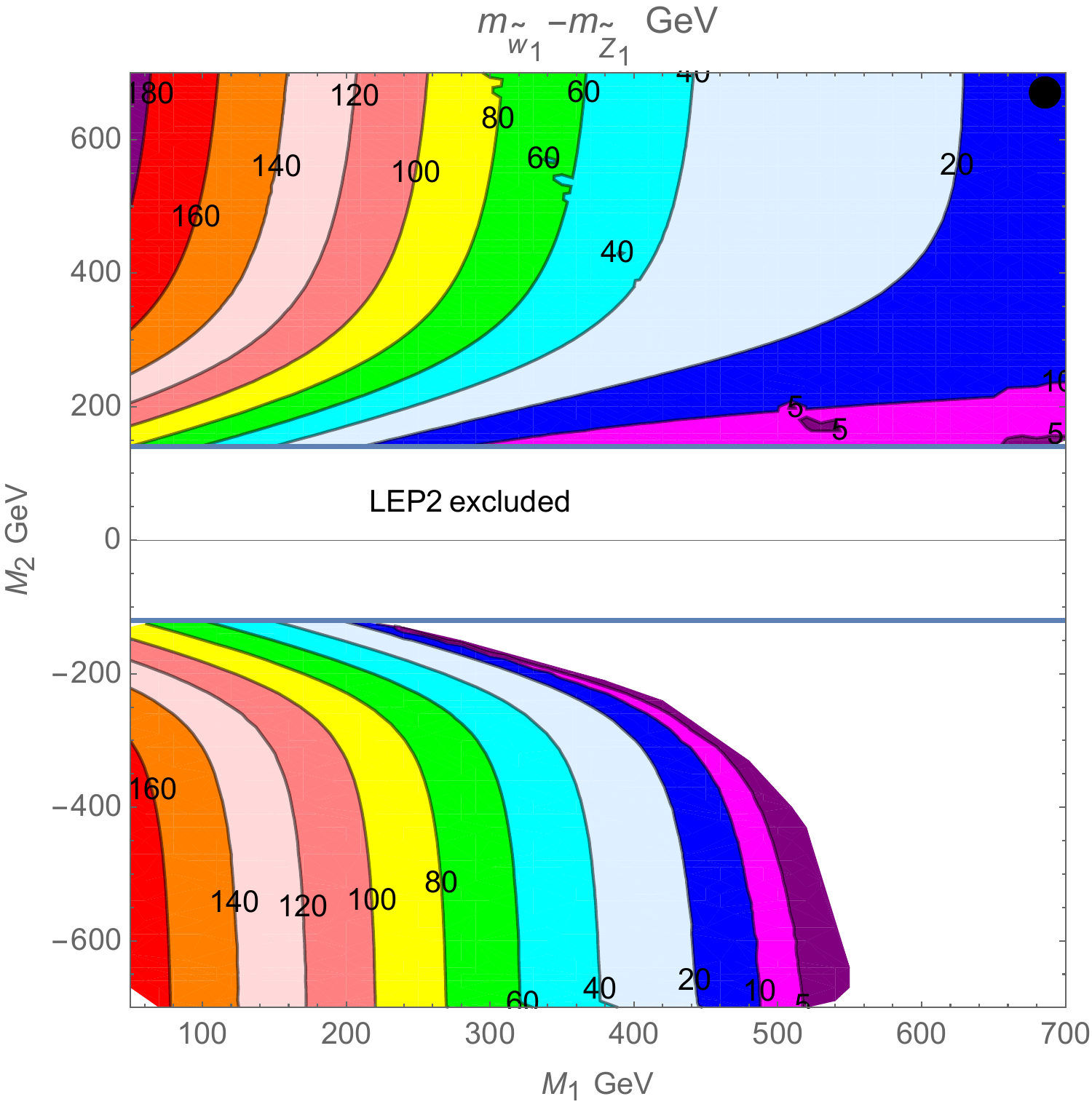}
\end{center}
\caption{The $m_{\tw_1}-m_{\tz_1}$ mass gap in the $M_1$ vs. $M_2$ plane 
for the RNS SUSY benchmark model.
\label{fig:dmwz}}
\end{figure}

In Fig. \ref{fig:dmwz}, we show the $\tw_1 -\tz_1$ mass gap in the $M_1$
vs. $M_2$ plane for the RNS benchmark model. The purple shaded region
has mass gaps between 10 and  20~GeV and corresponds to the bulk of the
higgsino-like LSP region along with the wino-like LSP region.  As
expected, the mass gap becomes small when $\tw_1$ and $\tz_1$ are both 
higgsino-like ($|\mu| \ll |M_{1,2}|$) or when these are both very
wino-like ($M_2\ll |\mu|$). It also becomes small along the boundary 
of the region in the lower half plane where the chargino becomes the
LSP. It is
mainly when one moves to small $M_1$, or large $|M_2|$ and moderate 
$M_1$, that this mass gap
exceeds 40-50~GeV, so that the daughter leptons from chargino decays are 
expected to be relatively hard.  

\begin{figure}[tbp]
\begin{center}
\includegraphics[height=0.4\textheight]{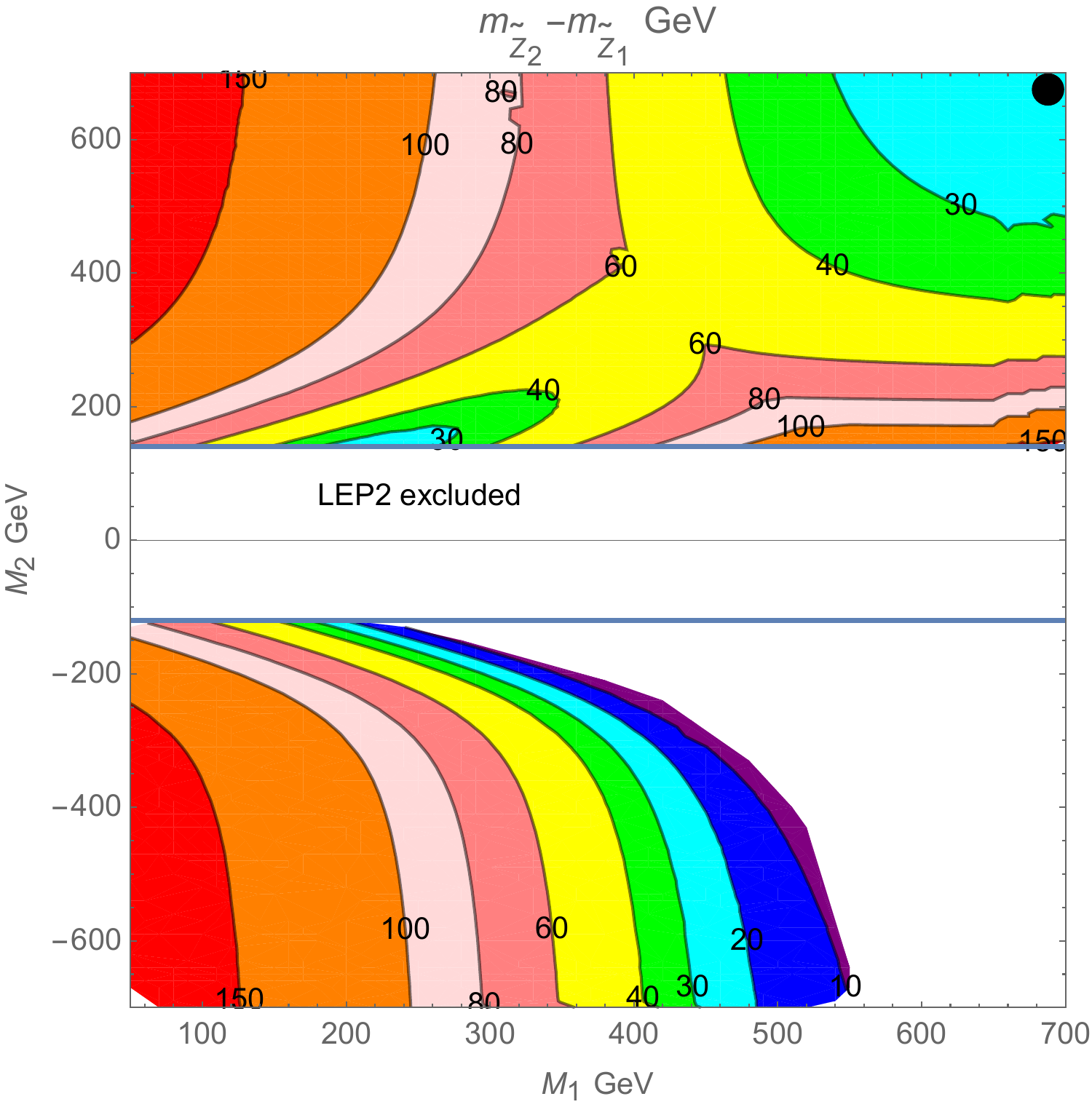}
\end{center}
\caption{The $m_{\tz_2}-m_{\tz_1}$ mass gap in the 
$M_1$ vs. $M_2$ plane 
for the RNS SUSY benchmark model.
\label{fig:dmz}}
\end{figure}

The $\tz_2 -\tz_1$ mass gap is shown in the $M_1$ vs. $M_2$ plane in
Fig. \ref{fig:dmz}.  Typically the smallest mass gap occurs
when we have a higgsino-like LSP as in the case of gaugino mass
unification in the upper right part of the plane, or in the region where $M_1$
and $|M_2|$ are both much larger than $\mu$ ($M_2< 0$). A small (purple) mass
region also occurs when $M_1\sim |2M_2|$ (lower half plane) so that
the weak scale values of $M_1$ and $|M_2|$ become comparable upon
renormalization group evolution: {\it i.e.} the bino and wino become
nearly degenerate, but the states remain nearly pure winos and binos
because of opposite signs of their mass terms.  
In this very low $\tz_2 -\tz_1$ mass gap region
one might expect enhanced bino-wino
co-annihilation(BWCA) in the early universe \cite{bwca}.
Note that the $\tz_2-\tz_1$
mass gap in especially the upper half plane, exceeds 50~GeV for a
large swath of the plane, and is larger than $M_Z$ and even $m_h$ 
over a substantial part. This should make for interesting signals
at LHC13 via the multilepton,  $WZ$ and $Wh$ plus $\eslt$ channels at
LHC13.  
It is also noteworthy that a region exists where {\it both} 
the $\tw_1-\tz_1$ and $\tz_2-\tz_1$ mass gaps fall below 10~GeV. 
This occurs in the narrow crescent at large $M_1$ in the lower half plane.
This region might be challenging even at the ILC if the 
heavier charginos and neutralinos are kinematically inaccessible.
In this case, techniques using initial state photon radiation might be 
required \cite{list}.

We note that while we have focussed on the mass gap between the lighter
charginos and neutralinos and the $\tz_1$, there is a substantial region of
the parameter space of natural SUSY models where signals from the
heavier charginos and neutralinos should be accessible at LHC13.
Because CMS and ATLAS LHC searches \cite{cms_ino,atlas_ino} tend to employ
hard cuts, it is entirely possible that signals from the heavy states
(assuming these are within the LHC13 reach) reveal themselves more
easily than signals for the lighter states.

\begin{figure}[tbp]
\begin{center}
\includegraphics[height=0.4\textheight]{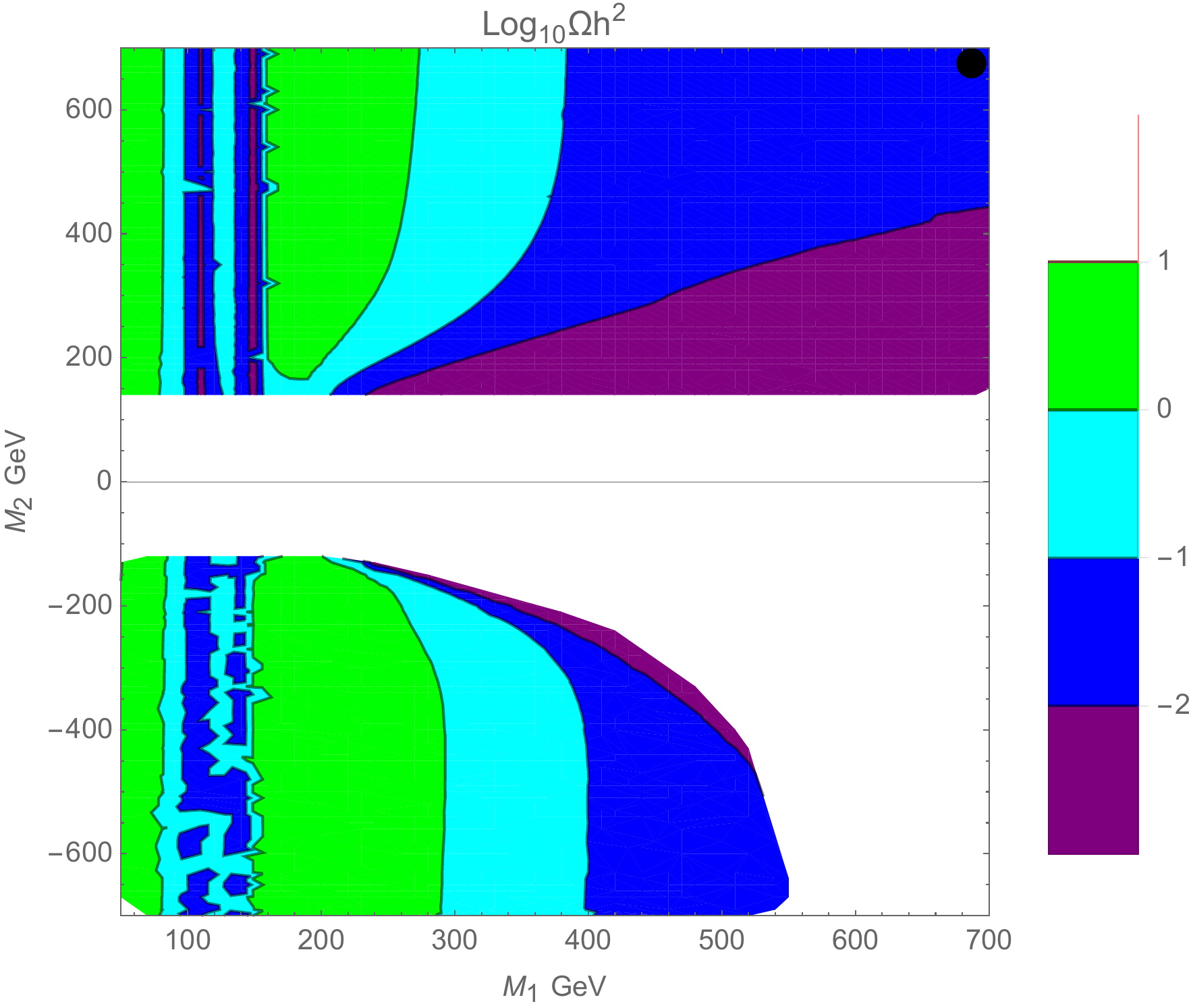}
\end{center}
\caption{Thermally-produced neutralino relic abundance in the 
$M_1$ vs. $M_2$ plane 
for the RNS SUSY benchmark model.
\label{fig:Oh2}}
\end{figure}

The thermally-produced neutralino relic density is shown in
Fig. \ref{fig:Oh2}.  The regions with very low relic density
$\Omega_{\tz_1}h^2\alt 0.01$ are 1. the wino-like LSP region along with 2. the
BWCA strip in the lower half plane where $m_{\tw_1}\simeq m_{\tz_1}$
and 3. the resonance annihilation regions where $2m_{\tz_1}\sim M_Z$ or
$m_h$ (the vertical strips at low $M_1$). The thermal relic density is
also below its observed value in the higgsino region or in parts of the
mixed bino-higgino LSP region. In these regions, we will need either
additional dark matter particles or non-thermal production of
neutralinos to match the measured value of cold dark matter relic density. 
The boundary of the light- and dark-blue shaded region is where we have a
well-tempered neutralino whose thermal neutralino relic density can
saturate the cold dark matter. In the light-blue and green-shaded parts
of the plane (deep in the bino LSP and away from the $Z$ and $h$
resonances) the relic density of neutralinos must be diluted by entropy
production late in the history of the Universe or else the 
$\tz_1$ must be made to decay either via $R$-parity violating interactions 
or decay to an alternative LSP ({\it e.g.} an axino).

We do not show the dark matter detection cross sections in this plane,
partly because for the most part we do not expect that these will
unambiguously constrain the parameter regions  for
reasons that we discussed earlier regarding the assumed local 
density of WIMPs.
%

\section{Conclusions:} 
\label{sec:conclude}

Supersymmetric models with radiatively driven naturalness are especially
interesting since they allow for $M_Z,\ m_h\sim 100$ GeV whilst
sparticles other than higgsinos can naturally be at the multi-TeV scale.
Such spectra seem to be required by reconciling naturalness with 
LHC8 sparticle search constraints
and with the measured value of the Higgs boson mass \cite{atlas_h,cms_h}. 
Most previous analyses have examined RNS models in
the context of gaugino mass unification. In that case, the LSP is
expected to be higgsino-like and constitute only a portion of the dark
matter while axions could make up the remainder.  The light higgsinos
required by naturalness can evade LHC searches because of
their compressed spectrum: higgsino decays release only small visible
energy, so that their production remains hidden under Standard Model
backgrounds.

These results follow from requiring {\it both} naturalness and gaugino mass
unification. We regard naturalness to be one of the main motivations for
supersymmetry.  In contrast, while gaugino mass unification is highly
motivated by the simplest GUT models, it is easy to construct GUTs with
non-universal gaugino masses at no cost to naturalness. Gaugino mass
non-universality results
if vacuum expectation values of the auxiliary 
fields that spontaneously break supersymmetry
also break the GUT symmetry.  The main requirement from LHC searches is
that $M_3\sim m_{\tg}\agt 1.3$ TeV.  The values of bino  and
wino mass parameters are relatively unconstrained.  If their weak scale
values are similar to or less than $|\mu|$, then the LSP can be either
bino-like or wino-like (or a mixture) instead of just higgsino-like at no
cost to naturalness. In such a case, both the collider expectations and
dark matter/WIMP search expectations change in important ways.

We have shown that in the case of natural SUSY models with enhanced bino
LSP content, increased mass gaps $\tw_1-\tz_1$ and $\tz_2 -\tz_1$ are
expected on account of bino-higgsino mixing.  The harder decay products
of $\tw_1$ and $\tz_2$ lead to discernable effects such as the presence
of dilepton mass edges in LHC events, and perhaps additional light
electroweak -ino pair production processes at the ILC involving also
$\tz_3$ production.  In the wino-like LSP case, then only the
$\tz_2-\tz_1$ mass gap opens up, while $\tw_1-\tz_1$ gap becomes
tighter. This situation should be readily discernable at ILC, especially
with the availability of polarized beams. Of course, if $M_2$ and
$|\mu|$ both assume modest values, the heavier states $\tw_2$,
$\tz_{3,4}$ will also be accessible at the LHC, and electroweak
chargino and neutralino production will lead to a rich variety of
multilepton, $WZ$ and $Wh$ plus $\eslt$ events
that are already being searched for \cite{cms_ino,atlas_ino,atlas_hino},
and possibly also spectacular $W^\pm W^\pm +\eslt$ events without
additional jet activity.
In such a scenario, ILC would become both a higgsino and a wino/bino factory
and it should be possible to perform a detailed bottom-up study of the
electroweak-ino sector, assuming that all states are kinematically
accessible \cite{zerwas}.

Expectations for WIMP searches also change. In the case of a bino-like
LSP, we generally expect a larger thermal abundance of neutralino dark
matter. While it is possible to obtain a well-tempered neutralino that
saturates the observed cold dark matter relic density, the thermal
neutralino density is often too large in which case it needs to be diluted by
late-time entropy production or else allowed to decay.
As a result, the
neutralino contribution to the relic density today depends on the
(unknown) physics, leading to significant uncertainties in prediction
of rates for direct
detection searches. While this makes it difficult to use experimental
bounds from LUX/XENON100\cite{lux} and other experiments to
unambiguously exclude portions of parameter space without a complete
model of particle physics and cosmology, these searches could lead to a discovery!

In the case of natural SUSY with a wino-like WIMP, then one expects an
even lower local abundance from thermally-produced neutralinos
as compared to the value for higgsino-like LSPs. 
The measured relic density must
then be made up by other (non-WIMP) relics of which axions may be the
most promising, or via WIMP production from late decays of heavy
particles. In view of the resulting uncertainty in the expectation for
local density of neutralino dark matter, we once again advocate using
caution when interpreting the absence of events in direct and indirect
dark matter searches to exclude ranges of model parameters. 
%

To sum up, in our view, supersymmetric GUTs remain the most attractive
solution to the naturalness problem plaguing the Standard Model and
light higgsinos are the most robust consequence of naturalness
considerations. If electroweak gaugino mass parameters happen to assume
modest values -- this is not required by naturalness but is completely
compatible with it -- there could be spectacular signals from
electroweak gaugino production at the LHC in multi-lepton+$\eslt$,
$WZ+\eslt$, $Wh+\eslt$ and $W^\pm W^\pm +\eslt$ channels.  Direct and
indirect searches for WIMPs could also reveal a signal even in the case
of a depleted local abundance of WIMPs.  
If natural supersymmetry is realized with fortuituously low gaugino masses,
then prospects for SUSY discovery at LHC13 will be vastly improved
since signals from {\em several} chargino and/or neutralino
states might also be observable.
Production of light electoweak -ino states at ILC -- 
as required by naturalness\cite{ilc} --
remains true but with even richer prospects since both gauginos and higgsinos
could be kinematically accessible.

\section*{Acknowledgments}

We thank A. Mustafayev for checking several calculations and for
pointing out an error in the first version of the text.
This work was supported in part by the US Department of Energy, Office of High
Energy Physics. HB would like to thank the William I. Fine 
Institute for Theoretical Physics (FTPI) 
at the University of Minnesota for hospitality
while this work was completed.

%

%
\end{document}